\documentclass[a4paper,11pt]{article}
\usepackage{jcappub} 
\usepackage{adjustbox}
\usepackage{lineno}

\arxivnumber{1234.56789} 
\title{The Einstein Gap: an unrecognised strong lensing feature in weak lensing mass density profiles}

\author[a,b]{David Crespo}
\author[a,b,1]{Joaqu{\'i}n Gonz{\'a}lez-Nuevo\note{Corresponding author.}}
\author[a,b]{Laura Bonavera}
\author[c,b]{Marcos M. Cueli}
\author[d]{Hu Zou}
\author[a,b]{Rebeca Fern{\'a}ndez-Fern{\'a}ndez}
\author[a,b]{Jose Alberto Cano}
\author[c,b]{Jose M. Casas}
\affiliation[a]{Departamento de Fisica, Universidad de Oviedo, C. Leopoldo Calvo Sotelo 18, 33007 Oviedo, Spain}
\affiliation[b]{Instituto Universitario de Ciencias y Tecnologías Espaciales de Asturias (ICTEA), C. Independencia 13, 33004 Oviedo, Spain}
\affiliation[c]{Departamento de Matemáticas, Universidad de Oviedo, C. Leopoldo Calvo Sotelo 18, 33007 Oviedo, Spain}
\affiliation[d]{National Astronomical Observatories, Chinese Academy of Sciences, A20 Datun Road, Chaoyang District, Beijing 100012, People’s Republic of China}

\emailAdd{gnuevo@uniovi.es}

\abstract{Magnification bias cross-correlations between galaxy clusters and high-redshift submillimetre galaxies (SMGs) have recently revealed a characteristic signal deficit at intermediate angular scales ($\sim 10''$, $\sim$50~kpc), termed the \textit{Einstein Gap}, that appears consistently across all lens types and cannot be explained by any single theoretical mass density profile. In this work we identify for the first time the most plausible physical origin of this feature through a combination of observational tests and improved lensing simulations. Using the 540,432galaxy clusters from the Zou et al. (2021) catalogue to characterise the satellite population, and the 9,051 clusters overlapping the H-ATLAS footprint as the lens sample, we show that the gap is detected at high statistical significance ($N_\sigma\simeq13$ for the full sample, and higher in the highest-mass subsamples), accounting for the full covariance of the stacked profile. Then, we demonstrate that the \textit{Einstein Gap} is robust across three independent estimators of the angular cross-correlation function, strengthens as the positional smoothing is reduced, and scales systematically with the bright central galaxy stellar mass. Finally, we show that the projected radial distribution of satellite galaxies, $\Sigma_{\rm sat}$, is smooth and continuous at all angular scales with no depression corresponding to the gap, disfavouring both a statistical artifact and a genuine absence of mass as its origin. An improved magnification bias simulator incorporating full ray-tracing via the lens equation and a NFW+Sersic mass density profile reproduces the observed signal deficit for an average cluster halo mass of $\sim 5\times10^{14}\,M_\odot$, strongly supporting that the \textit{Einstein Gap} is a strong lensing feature: background sources within the Einstein radius are displaced outward, creating a characteristic empty annulus whose angular size scales as $\theta_E \propto M_{\rm H}^{2/3}$. This feature is consistent with features visible in previously published weak-lensing profiles that were not, however, interpreted as a strong-lensing signature. Its identification opens new possibilities for constraining halo masses and concentrations from magnification bias measurements alone.}

\begin{document}
\maketitle
\flushbottom

\section{Introduction}
\label{sec:intro}

When a foreground object, also known as a lens, produces a gravitational 
lensing effect, the light rays from background sources are magnified and 
the area of the surrounding sky region is stretched. These effects modify 
the number counts of the lensed objects, resulting in an excess or deficit 
of background sources at a given flux density limit, which is primarily 
related to the logarithmic slope of the integrated number counts, $\beta$. 
This gravitational lensing effect is known as magnification bias 
\citep{SCH92}. For this excess of background sources to occur, a value of 
$\beta>1$ is required \citep{BON22}, and the sensitivity of magnification 
bias increases significantly for cases with very steep source number counts, 
$\beta \gtrsim 3$, making the weak lensing effect more easily detectable. This weak 
gravitational lensing provides direct evidence of dark matter around the 
lenses and results in a non-zero angular cross-correlation function (CCF) 
between lens and source samples even with non-overlapping redshift 
distributions, which can only be explained by magnification bias 
\citep{SCR05, MEN10, HIL13, Bar01}.

The ideal sources for such CCF studies are SubMillimetre Galaxies (SMGs), 
as shown in \cite{GON14,GON17}. The unprecedented data provided by Herschel 
\citep{EAL10} were of great significance for the physical understanding of 
SMGs in terms of area and sensitivity. Those galaxies have a high redshift 
distribution ($z>1$), very steep source number counts ($\beta \gtrsim 3$), and emit the bulk of their radiation in the submillimetre due to dust obscuration, making them virtually invisible in the optical.

In the wider context of cosmic environments, galaxy clusters are massive bound systems usually placed at the intersections of filamentary structures, allowing us to study the large-scale structure of the Universe. They have played a crucial role in different areas, such as cosmological studies \citep{ALL11}, galaxy evolution \citep{DRE80, BUT78, BUT84, GOT03, BEL16, MUL18}, and the study of high-redshift lensed galaxies \citep{BLA99, SCH17, FOX22}. The correlation between galaxy clusters and selected background objects has been employed to 
investigate potential lensing effects \citep{MYE03}, including CCF 
measurements using SMGs as a background sample to estimate the mass and 
concentration of galaxy clusters \citep[][hereafter FER22]{FER22}.

Stacking techniques are commonly used when the signal to be detected is weak 
but occurs frequently, as is the case with weak lensing. Individual weak 
lensing events are impossible to detect, but by co-adding the emission from 
many such events, stacking enhances the overall signal while reducing 
background noise, allowing for a robust statistical analysis. This method 
has been applied in various contexts, including the detection of the faint 
Sachs--Wolfe effect in Planck data \citep{Pla14, Pla16b}, the study of 
polarised signals from radio and infrared sources in the NRAO VLA Sky Survey 
(NVSS) and Planck \citep{Stil14, Bon17a, Bon17b}, and deriving the mean 
spectral energy distribution of optically selected quasars \citep{Bia19}. 
It has also been used to identify weak gravitational lensing in the Planck 
lensing convergence map \citep{Bia18}, to explore star formation in the 
dense environments of $z\sim1$ lensing halos \citep{Wel16}, and galaxy-galaxy lensing \citep{MAR15, WAN24}. More recently, FER22 applied stacking to measure the CCF between galaxy clusters of varying richness and SMGs. 
Additionally, \cite{UME16} employed stacking techniques to estimate the 
average surface mass density profile of an X-ray-selected subsample of 
galaxy clusters by combining their individual profiles.

In previous studies \citep[][hereafter CRE22 and CRE24]{CRE22, CRE24}, we demonstrated that the magnification bias effect allows the extraction of mass density profiles for various types of lenses (QSOs, galaxies, and clusters). The stacking techniques with the enhanced positional accuracy of the SMGs provided by NASA's Wide-field Infrared Survey Explorer (WISE) All-Sky Data Release \citep[][]{WIS10}, with a positional uncertainty of $\sigma_{pos} = 0.3''$, enabled CRE24 to achieve measurements probing physical scales below $\sim$5 kpc. Such central scales are precisely where conventional shear analyses face serious systematic limitations: intrinsic alignments of galaxies near the lens \citep{MAN06}, bright central galaxy (BCG) miscentering \citep{JOH07}, and the contribution of satellite subhaloes \citep{YAN06, LI09} that become entangled with the smooth halo signal when individual galaxy lensing profiles are stacked. The magnification bias approach, being based on number counts rather than shape measurements, is largely immune to these systematics, making it particularly well suited to probe the innermost regions of mass density profiles.

The CRE24 measurements reveal that the profiles show an excess of signal in the central region and a marked lack of signal at intermediate angular scales ($\sim 10''$). This feature, which we term the \textit{Einstein Gap}, 
appears consistently across all lens types and catalogues studied in CRE24, and cannot be explained by a single theoretical mass density profile. Intriguingly, a signal deficit at similar physical scales was noted but not 
further investigated in the highly cited galaxy-galaxy weak lensing study of \cite{MAN06}, and in other studies using different techniques and samples 
\citep{HUD98,HOE03,HOE05,GAV07,JOH07,LI09,Man09,SHE09,NEW13,BAU14,HUD15,ZU15, UME16,UIT16,VEL17,LUO17,LUO22,RAN22,MIS23,XU24}, suggesting that the \textit{Einstein Gap} may be a general feature of gravitational lensing mass density profiles rather than a peculiarity of our methodology or source sample.

The radial distribution of satellite galaxies also provides an independent and complementary probe of the cluster mass distribution. The observed distribution of cluster galaxies around the virial radius is known to follow the Navarro-Frenk-White \citep[NFW,][]{NAV96} profile predicted for dark matter \citep[e.g.][]{WAT12, TAL12, GUO13}, though with lower concentrations \citep{NAG05, van2015evidence}. Several studies show that this distribution depends on the stellar mass of the central galaxy, as evidenced by observational works \citep{WAN14, GU22} and supported by simulations \citep{TOR97, VAN05}. Moreover, since satellites can produce their own lensing effects, they can enhance the observed magnification bias signal 
\citep{MEN20,MEN22,MEN23}, and comparing the satellite and lensing profiles allows us to disentangle contributions from the cluster halo, the central galaxy, and the satellite population. 

Despite its consistent appearance, the physical origin of the \textit{Einstein Gap} 
remained unexplained in CRE24. Three candidate explanations were left open: 
(i) a statistical artifact arising from insufficient numbers of lensing 
events; (ii) a genuine absence of mass at those angular scales; or (iii) a 
lensing effect not fully captured by the weak lensing approximation. 
The present work goes beyond CRE24 in four specific respects: 
\begin{enumerate}
    \item We identify satellite galaxies, via the extensive ZOU catalogue, as an independent, lensing-free tracer of the underlying mass distribution, and use it to exclude a genuine absence of mass as the origin of the gap.
    \item We develop a magnification bias simulator that, unlike the purely weak-lensing approach of CRE24, fully incorporates the lens equation via ray-tracing, allowing us to reproduce the gap as a direct consequence of strong lensing rather than merely characterise it.
    \item We demonstrate, for the first time, that the depth and angular position of the gap scale systematically with BCG stellar mass across independent subsamples, in the manner predicted for a strong-lensing feature \citep[$\theta_E \propto M_{\rm H}^{2/3}$;][]{LAP12}.
    \item We combine these results with a systematic set of observational robustness tests (Section~\ref{sec:method}) to discount statistical and methodological artefacts.
\end{enumerate}
Individually, some of these tests refine methodology already present in CRE24. However, taken together, they constitute the first identification of the physical mechanism responsible for the \textit{Einstein Gap}, rather than an incremental update of the measurement itself.

The organisation of the paper is as follows. Section~\ref{sec:data} describes the data. Section~\ref{sec:method} provides a detailed description of the measurement methodology, including the three-estimator comparison that establishes the robustness of the \textit{Einstein Gap}, the satellite number density profile analysis that strongly disfavours a mass deficit as its origin, and the mass-binned analysis that demonstrates its scaling with BCG stellar mass. Section~\ref{sec:eins} presents the improved magnification bias simulator, discounts the remaining alternative explanations, and provides compelling evidence that support the strong lensing origin of the \textit{Einstein Gap}. Section~\ref{sec:concl} summarises our conclusions and suggests avenues for future research. Theoretical details of the magnification bias formalism and the mass density profiles used in this work are provided in Appendix~\ref{ANX:profiles}. Appendix~\ref{ANX:Validation} presents additional robustness tests, including the validation of the analytical random signal estimation, the comparison of cross-matching strategies, and the dependence of the \textit{Einstein Gap} on the positional smoothing kernel. Finally, Appendix~\ref{ANX:figures} contains additional figures illustrating the miscentring effects in low-mass systems. Throughout the paper, the cosmological model adopted is a flat $\Lambda$ cold dark matter ($\Lambda$CDM) model with cosmological parameters estimated by \cite{PLA18_VI}: $\Omega_m = 0.31$, $\sigma_8 = 0.81$, and $h = H_0/100$ km\,s$^{-1}$\,Mpc$^{-1} = 0.67$.

\section{Data}
\label{sec:data}

\subsection{Membership samples}

To improve the preliminary test on the radial distribution of cluster satellites with respect to the position of the BCG conducted in CRE24, we use the extensive cluster catalogue from \cite{ZOU21, ZOU22} (hereafter ZOU). The galaxy cluster catalogue was obtained from DESI Legacy Imaging Surveys \citep{DESI} and provides 540,432 galaxy clusters spanning an area of approximately 20,000 deg$^2$.  The detection of cluster positions in this catalogue is based on the measurement of local ($\Phi$) and background density ($\Phi_{bkg}$). Using this method, the peaks are identified as the galaxies with the highest local density, where $\Phi > 4\Phi_{bkg}$. As an alternative to using the peak as the cluster centre (PEAK centre), the BCG position can also be taken as the centre, identified as the brightest galaxy within 0.5 Mpc of the peak (BCG centre). The cluster members are then determined by grouping galaxies around these centres \citep[more details in][]{ZOU21,ZOU22}. It should be noted that this cluster centre is not always the BCG that is usually close to the densest region. This distinction is particularly important and will be further addressed in later sections. According to this method, the typical search radius for identifying members and defining a cluster is 1 Mpc. The catalogue does not provide any information on the integrity and contamination rate of the cluster members. In our case, we are not interested in distances greater than 1 Mpc. Therefore, for the purposes of our analysis, both issues can be ignored. Thus, from the 540,432 clusters in the catalogue, we collected a total of 12,046,450 satellite galaxies and their positions relative to their corresponding BCG.\\

\begin{figure}
  \centering
    \includegraphics[width=0.9\linewidth]{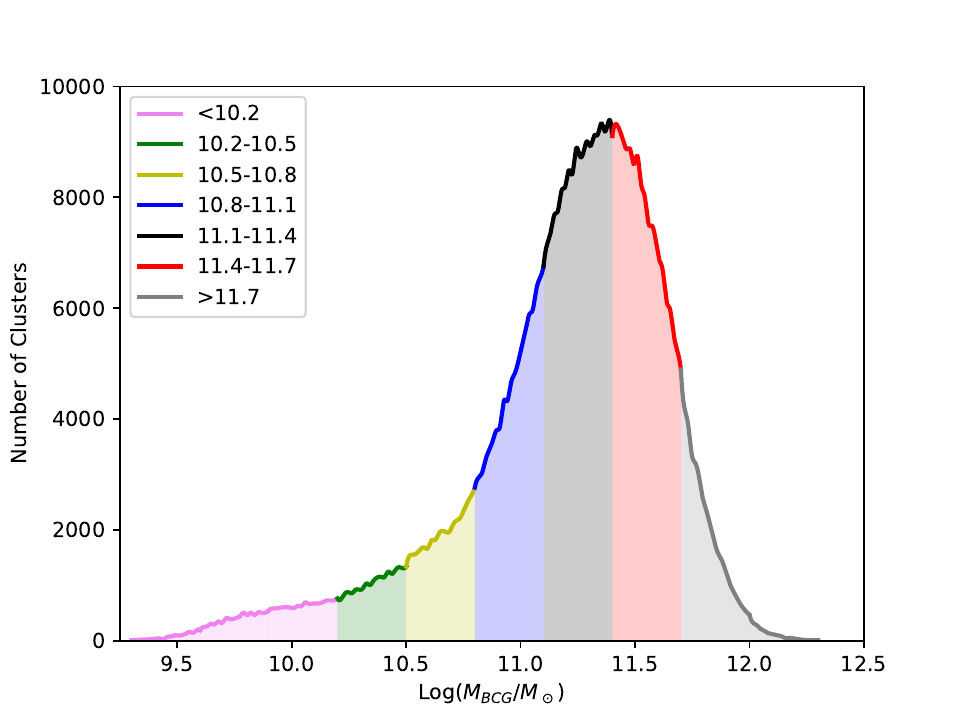}
  \caption{Mass distribution of BCGs, divided into mass bins. The colour-coded regions represent different stellar mass ranges, in $\log(M_{BCG}/M_\odot$), from less than $10.2$ (pink) to greater than $11.7$ (grey).}
  \label{fig:mass_hist}
\end{figure}

Moreover, ZOU gives us the value of the stellar mass for each BCG. Therefore, we can construct a mass distribution of the BCG, as seen in Fig. \ref{fig:mass_hist}. Using this information, we can analyse the behaviour of the signal loss observed in CRE24 for seven different mass ranges, whose characteristics can be found in Table \ref{tableMass}.

\begin{table*}
\centering
\begin{adjustbox}{width=\textwidth}
\begin{tabular}{ccccccccc}
\hline
    & BIN 1 & BIN 2 & BIN 3 & BIN 4 & BIN 5 & BIN 6 & BIN 7 \\
    & \textless{}10.2 & 10.2 - 10.5 & 10.5 - 10.8 & 10.8 - 11.1 & 11.1 - 11.4 & 11.4 - 11.7 & 11.7 \\ \hline
\multicolumn{1}{c}{$ \langle {\rm M}_{\rm BCG} \rangle$}  & 9.92           & 10.37      & 10.67      & 10.97      & 11.26      & 11.53      & 11.81             \\ 
\multicolumn{1}{c}{\# Satellites} & 372,746         & 373,758      & 707,974 & 1,746,667& 3,611,576 & 3,816,625 & 1,413,393             \\ 
\multicolumn{1}{c}{\# BCG} & 22,659& 21,226 & 39,379 & 92,304 & 169,778 & 151,732 & 43,188 \\ 
\multicolumn{1}{c}{\# BCG lenses} & 324         & 318      & 611      & 1,556      & 2,842     & 2,653      & 747             \\
\multicolumn{1}{c}{$\langle z\rangle$} & $0.63^{+0.17}_{-0.38}$           & $0.54^{+0.25}_{-0.33}$    & $0.56^{+0.23}_{-0.35}$      & $0.50^{+0.27}_{-0.34}$      & $0.48^{+0.27}_{-0.31}$    & $0.46^{+0.29}_{-0.30}$      & $0.48^{+0.27}_{-0.30}$             \\ 
\\
PEAK as centre:\\
M$_{\rm NFW}$       & $13.63^{+0.16}_{-0.21}$    & $13.70^{+0.04}_{-0.04}$   & $13.58^{+0.18}_{-0.31}$    & $14.12^{+0.06}_{-0.07}$    & $14.10^{+0.04}_{-0.05}$    & $13.88^{+0.03}_{-0.04}$    & $14.11^{+0.02}_{-0.03}$             \\
C$_{\rm NFW}$       & $0.04^{+0.01}_{-0.01}$   & $0.08^{+0.01}_{-0.01}$    & $0.13^{+0.01}_{-0.02}$     & $0.38^{+0.01}_{-0.01}$    & $0.72^{+0.01}_{-0.01}$    & $0.99^{+0.01}_{-0.02}$    & $1.19^{+0.01}_{-0.01}$               \\ 
BCG as centre:\\
M$_{\rm NFW}$       & $12.74^{+0.67}_{-1.25}$ & $13.63^{+0.27}_{-0.41}$ & $13.88^{+0.11}_{-0.13}$ & $13.38^{+0.18}_{-0.23}$ & $13.60^{+0.14}_{-0.18}$ & $14.02^{+0.01}_{-0.01}$ & $14.01^{+0.06}_{-0.07}$              \\
C$_{\rm NFW}$       & $0.03^{+0.01}_{-0.02}$  & $0.14^{+0.01}_{-0.02}$  & $0.14^{+0.01}_{-0.01}$  & $0.30^{+0.03}_{-0.04}$  & $0.61^{+0.04}_{-0.05}$  & $1.13^{+0.01}_{-0.01}$  & $1.17^{+0.01}_{-0.02}$               \\
\\
M$_{\rm SHMR}$       & 11.69 & 11.96 & 12.26 & 12.91 & 13.90 & 14.83 & 15.83 \\
\hline
\end{tabular}
\end{adjustbox}
\caption{Summary of the clusters sample analysis. From top to bottom: The BIN number and its range of the BCG logarithmic stellar mass in 
$M_\odot$, the mean BCG logarithmic stellar mass in $M_\odot$, the number of satellites, the number of BCGs, the number of BCGs potentially acting as lenses in the common area between both surveys and the mean redshift where the uncertainty represents the 95\% percentile range. Then, the logarithmic total mass in $M_\odot$ and concentration derived from the NFW fit using the PEAK and BCG positions as centres. Finally, the expected logarithmic total mass in $M_\odot$ from a traditional SHMR \citep{MOS13}.}
\label{tableMass}
\end{table*}

\subsection{Foreground sample}
To compare with the direct satellite number density profile, we will employ galaxy clusters as gravitational lenses, as done in previous works. These clusters were extracted from the same catalogue as the satellite analysis, which offers the advantage of spatially overlapping with the galaxies detected in the Herschel Astrophysical Terahertz Large Area Survey (H-ATLAS) data. 

\begin{figure}[h]
  \centering
    \includegraphics[width=\linewidth]{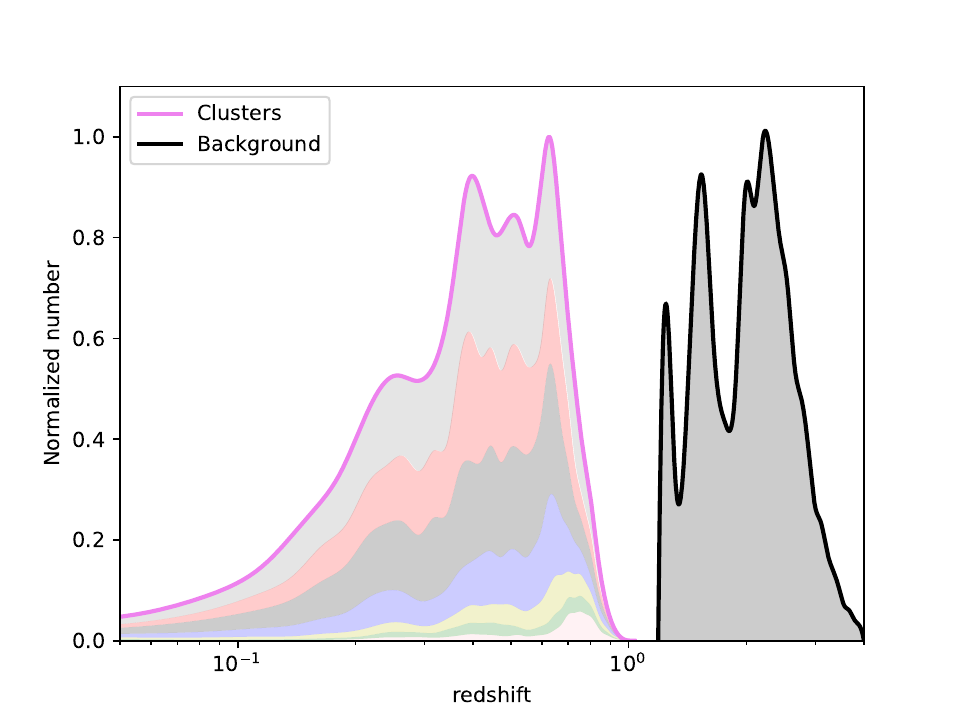}
  \caption{Redshift distribution of background SMGs selected in WISE (in black) and foreground clusters (in pink). Within the clusters, the contributions by mass ranges can be observed, following the same colour scheme as Figure \ref{fig:mass_hist}, arranged from bottom to top starting with BIN 1, for masses below $10^{10.2}M_\odot$ , up to BIN 7, for masses above $10^{11.7}M_\odot$. }
  \label{fig:z_hist}
\end{figure}

To minimize the potential overlap and contamination between the foreground and background samples in terms of redshift, we restricted our selection to clusters with redshift $z<1$ (see Fig. \ref{fig:z_hist}). As a result, we identified over 9,000 clusters, with a mean redshift of 
$\langle z\rangle=0.50^{+0.24}_{-0.30}$, where the uncertainty represents the 95\% percentile range.

To perform a more direct comparison between the results obtained from the satellite distribution and those obtained through lensing, we applied the same mass binning as used in Fig. \ref{fig:mass_hist}. The number of lenses in each bin is specified in Table \ref{tableMass}.

\subsection{Background sample}
In previous studies \citep{GON17,GON21, BON19, BON20,FER22, CRE22, FER24} the SMG background samples were based on detections from Herschel Astrophysical Terahertz Large Area Survey \citep[H-ATLAS, ][]{PIL10}. These observations were conducted using the Herschel Space Observatory and cover an area of approximately 610 square degrees across five distinct fields. Three of these fields, known as the Galaxy and Mass Assembly (GAMA) fields, are located on the celestial equator at 9, 12, and 14.5 hours (G09, G12, and G15). The remaining fields, referred to as the North and South Galactic Poles (NGP and SGP), span areas of 180.1 and 317.6 square degrees, respectively.

In order to improve the positional error, following the approach of CRE24, the current sample is based on additional data from WISE \citep{WIS10}. While WISE does not directly detect SMGs, its data, when combined with the two instruments used in H-ATLAS, namely the Photodetector Array Camera and Spectrometer (PACS) and the Spectral and Photometric Imaging Receiver (SPIRE), significantly improves the positional accuracy of SMGs. A cross-match between both catalogues was performed to obtain a sample of SMGs with a positional accuracy well described by a Gaussian distribution with $\sigma_{\rm pos} = 0.3''$ \citep{WIS10}, improving on the $\sigma_{\rm pos} = 2.4''$ provided by the H-ATLAS catalogue positions \citep{VAL16, BOU16, MAD18}. The details of the cross-matching procedure are given in Appendix \ref{ANX:xmatch}\\

To avoid overlap between the redshift distributions of the sources and the lenses, the redshift range for the SMGs was restricted to $1.2 \leq z \leq 4$. Their redshift distribution is shown in Fig. \ref{fig:z_hist} in black.

\section{Measurements}
\label{sec:method}
\subsection{Magnification Bias CCF}
\label{sec:stack}

Instead of the traditional CCF, employed in works like \cite{GON14,GON17}, we use a statistical method called stacking, which is particularly useful in cases like ours, where the noise may be comparable to the signal. It involves combining multiple weak lensing events to strengthen the signal, which would otherwise be too weak to be detected in individual cases.

In previous works, this method has been used to study the CCF signal resulting from magnification bias between background SMGs and objects acting as lenses. In \cite{BON19}, QSOs were used as lenses, while in FER22, galaxy clusters were employed. The study by CRE22 re-analysed this effect using both QSOs and galaxies. Finally, CRE24 conducted a comprehensive study involving galaxies, QSOs, and clusters simultaneously. This was made possible due to the improved positional accuracy provided by the WISE catalogue, allowing for more precise measurements and the ability to probe smaller distances approaching a few kiloparsec scales.

The measuring procedure is similar to that used in previous studies, where a search radius is defined within a circular area centred on the lens position to identify background sources. This approach is similar to the traditional two-point cross-correlation estimator but offers the added benefit of accounting for positional errors, and the identification of foreground-background pairs. The positions of the pairs are recorded on a square map with a specific pixel size and number of pixels, determined by the search radius and the desired angular resolution, which is limited by the positional accuracy of the catalogues. To account for the positional accuracy of the catalogues, a Gaussian convolution kernel with a standard deviation corresponding to the positional accuracy of the background sample was applied to the stached maps. For this study, we focus mainly on the outer region of the CCF, and therefore, we can maintain the more conservative $\sigma_{\rm pos}=2.4''$ that produces smoother, less noisy measurements. In this case, we require only a search radius of $100''$ and a pixel size of $0.5''$, resulting in a stacked map of $400\times400$ pixels centred on the lens position, containing the paired background sources. The maps obtained for all the lenses are then aggregated and normalized to the total number of lenses, producing the final stacked map.

To estimate the CCF from the maps, we draw a finite set of concentric circles centred at the central position of the images. The radii increase logarithmically in steps of 0.05, starting with $1''$ (the first measurements are limited by the pixel size). This defines one initial circle and a set of rings. The pixel values in each circular annulus are added up as the data-data term (DD) using the typical correlation-analysis nomenclature. Instead of using a random map to generate the Random-Random (RR), as in \cite{BON19}, FER22 and CRE22, we computed the RR term theoretically from the annulus considering a constant surface density (CRE24). The details and validation of this analytical RR estimation are described in Appendix~\ref{ANX:Validation}. The standard estimator \citep{Dav83} is then calculated as:
\begin{equation}
    \tilde{w}_x=\frac{DD}{RR}-1
    \label{ec:ccf}
\end{equation}

To compute the CCF uncertainty, the first four rings were divided into sections of equal area. However, the subsequent rings had enough pixels to allow for division into 15 sections of equal area. Following the approach described in \cite{BON19}, a Jackknife method was applied within each annulus to estimate the uncertainties for DD. These uncertainties were then propagated using Eq. \ref{ec:ccf}  to determine the uncertainty of the overall signal. 

\begin{figure}[h]
 \centering
\includegraphics[width=\textwidth]{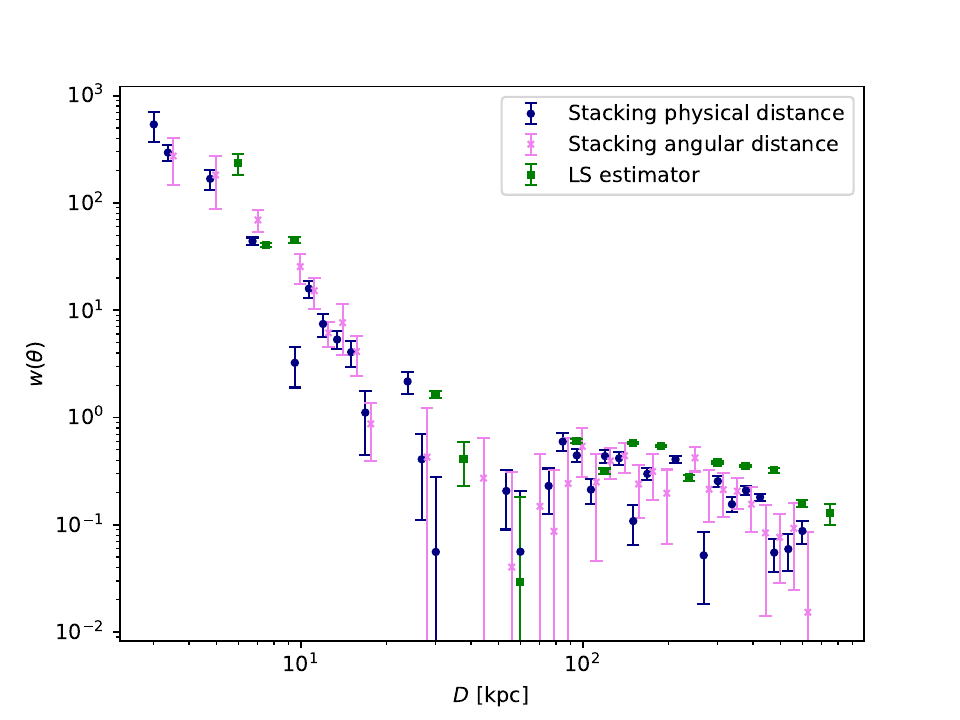}
\caption{Comparison of three methodologies for estimating the CCF between ZOU clusters (lenses) and H-ATLAS/WISE sources (background): the traditional Landy-Szalay estimator (green), stacking in angular distance with 
$\sigma_{\rm pos}=0.3''$ smoothing (light magenta), and stacking in physical distance with equivalent smoothing (dark blue). All three approaches consistently recover the same main features: a strong central signal at $\lesssim 20$~kpc, the \textit{Einstein Gap} at $\sim 50$~kpc, and a flat outer profile, demonstrating the robustness of these features to the choice of estimator or distance convention. Differences between methods at the smallest and largest scales are discussed in the text.}
\label{fig:methods}
\end{figure}

Before proceeding to the real analysis, we assess the robustness of our measurements by comparing different methodologies to estimate the cross-correlation between the ZOU clusters as lenses and the H-ATLAS sources as the background sample (with the improved positions from WISE; see Appendix~\ref{ANX:xmatch} for more details). In Fig.~\ref{fig:methods} we compare the CCF estimated using three different approaches. The CCF from the traditional Landy-Szalay estimator \citep[][and references therein]{GON26} is shown in green; no smoothing step was applied, as the positional uncertainty is difficult to model within this kind of estimator. The results from the stacking method using angular distances, as described above, are shown in light magenta; for this comparison, the smoothing was performed adopting the WISE positional uncertainty, $\sigma_{\rm pos}=0.3''$, to make the comparison as close as possible to the traditional estimator. Finally, the CCF estimated via stacking using physical distances, translated from angular separations accounting for each individual lens redshift, is shown in dark blue. The positional uncertainty is not well defined in this case, but we applied a smoothing step analogous to that of the angular stacking method for the sake of comparison.

All three approaches are qualitatively equivalent and reproduce the same main features already identified in CRE24: a strong central signal at $\lesssim 20$~kpc, the \textit{Einstein Gap} at $\sim 50$~kpc, and a flat outer profile. At the smallest scales, the stacking method outperforms the traditional estimator, which yields many negative measurements owing to the large number of pairs required for robust estimates. At intermediate separations, all three methods give fully compatible results supporting the \textit{Einstein Gap}, although the traditional estimator achieves smaller uncertainties. At the largest scales, $\gtrsim500$~kpc, the stacking measurements become less reliable as we approach the maximum search radius and the field density begins to fluctuate around the mean value assumed by each method. These systematics are better handled by the traditional Landy-Szalay estimator, which is therefore the standard choice for cosmological analyses requiring measurements at even larger angular separations \citep[see][and references therein]{GON26}.

Comparing the two stacking variants, stacking in physical distance reduces the uncertainty of individual bins but produces many measurements with anomalously low values, likely due to an insufficient number of pairs at certain separations. Stacking in angular distance yields slightly larger uncertainties but is more robust and consistent overall. Crucially, this comparison reinforces the robustness of the \textit{Einstein Gap} detection and clarifies why it is not smeared out when stacking in angular distance, even though a given angular separation could in principle map to a range of physical scales. The stacking technique is essential precisely because the probability of finding a lens--source pair at such separations is very low for any individual lens (typically $\sim$0.001, \cite{LAP12}). A statistically significant signal, or deficit, therefore only emerges from lenses near the peak of the redshift distribution, where the contribution of the majority of pairs is concentrated. For this reason, both stacking approaches are nearly equivalent, and stacking in angular distance, combined with a conversion to physical scales using the mean foreground redshift, provides a more robust result than stacking directly in physical distance. Accordingly, angular-distance stacking is adopted as our default method throughout the remainder of this paper.

\begin{figure}[h]
 \centering
\includegraphics[width=0.51\textwidth, height=0.47\textwidth]{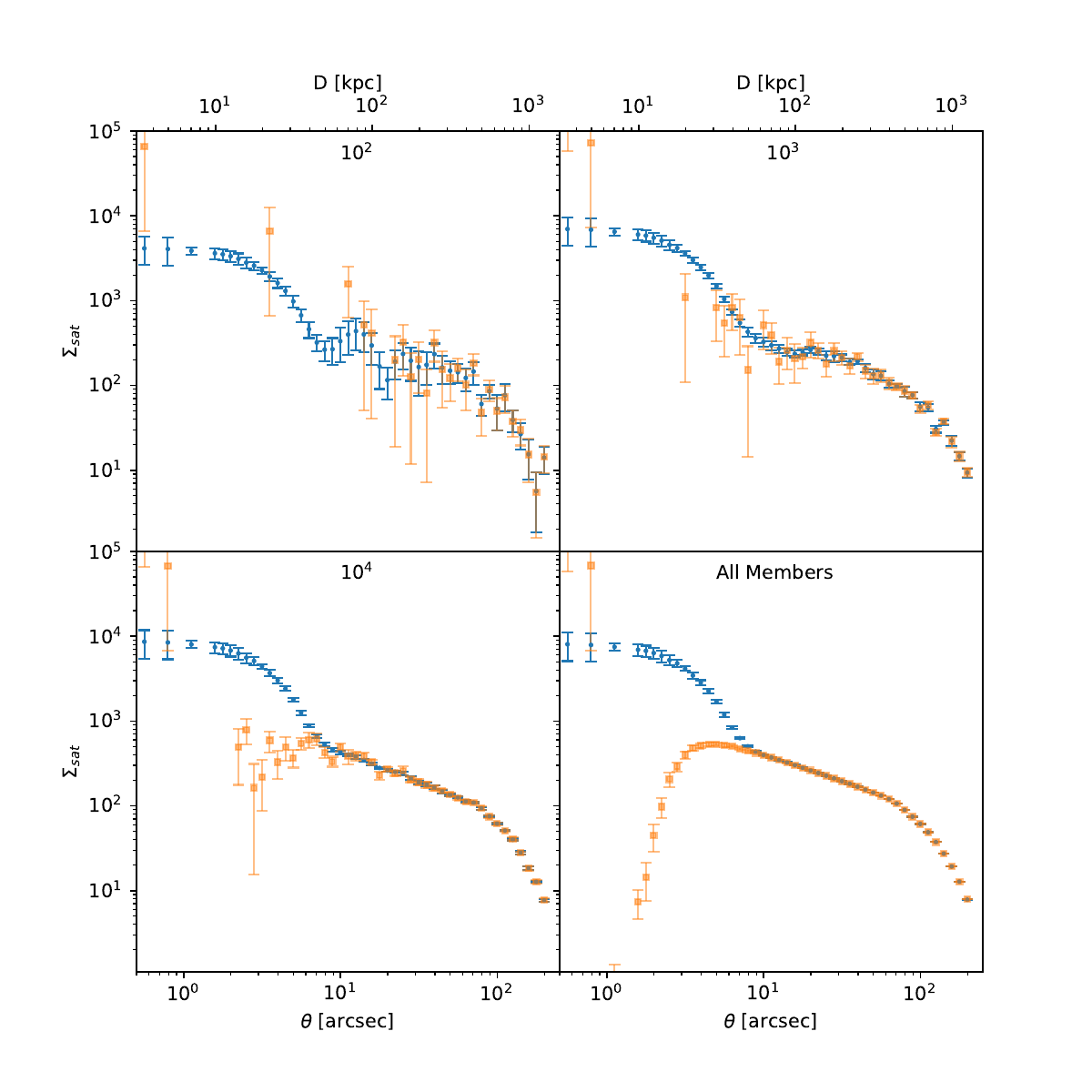}
\includegraphics[width=0.45\textwidth]{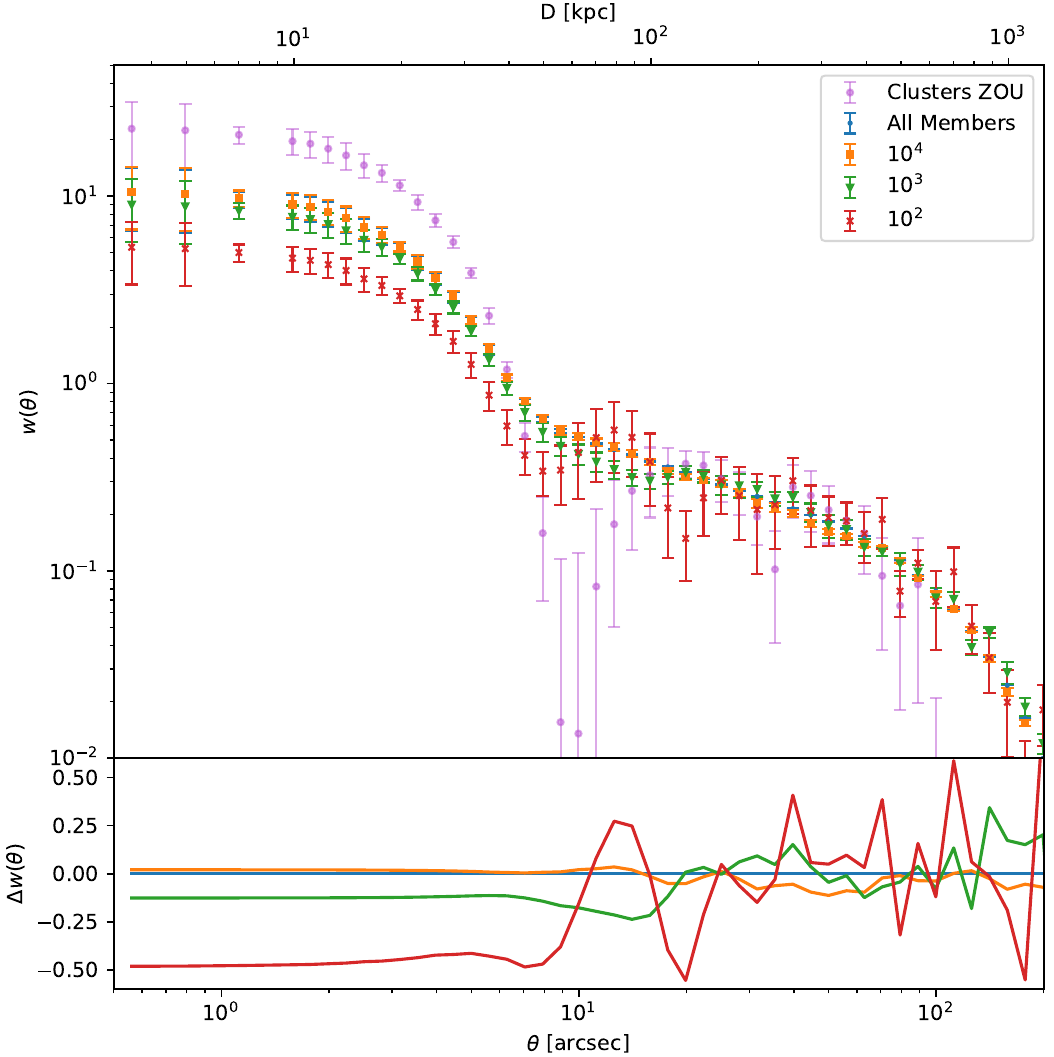}
\caption{\textit{Left}: Satellite number density profile derived by stacking satellite positions around the BCG without any positional uncertainty (orange points) and assuming  a $\sigma_{pos}=2.4''$ one (blue points). The number of satellites used is indicated inside each panel. \textit{Right}: Comparison between the lensing results (in magenta) of the clusters from the ZOU catalogue and the results of stacking the satellite positions (there is no lack of signal in any of the cases). The members are randomly selected, using 4 subsamples of satellites with sizes of $10^2$ (in red), $10^3$ (in green), $10^4$ (in orange), and the total set of satellites (in blue), approximately $1.2 \cdot 10^7$ (more information in the text).  The difference between using the total number of satellites and the other 3 cases is plotted in the bottom panel.}
\label{fig.zou_vs_members}
\end{figure}

Finally, our baseline CCF measurement for the ZOU sample, obtained using the more conservative smoothing of $\sigma_{\rm pos}=2.4''$, is shown in Fig.~\ref{fig.zou_vs_members} (light magenta points); this is a slightly smoother version of the profile discussed in this subsection.

\subsection{Satellite number density profile}
\label{sec:sat_ndp}

As discussed in the introduction, one of the candidate explanations for the \textit{Einstein Gap} is an actual deficit of mass at intermediate angular scales ($\sim$50~kpc), which would suppress the lensing signal independently of the methodology used. To test this hypothesis, we study the projected radial distribution of satellite galaxies around the BCG as a direct, though imperfect, tracer of the underlying matter distribution within the cluster halo. If the \textit{Einstein Gap} were caused by a genuine lack of mass at these scales, a corresponding depression should appear in the satellite number density profile, $\Sigma_{\rm sat}$, at similar physical separations.

The stacking method described in the previous subsection is particularly well suited for this purpose, and can be naturally extended to map the satellite distribution around the BCG position. Due to the large number of satellites, the search radius for this case can be $200''$, requiring the construction of a square map with a pixel size of $0.5''$ and $800\times800$ pixels centred at the BCG position. By applying this method to each of the clusters, adding up all the maps and normalising to the total number of clusters, we obtain a stacked map representing the average satellite distribution.

Using the sample of cluster members, we created four different subsamples: one consisting of the entire set of members, and the other three containing $10^2$, $10^3$ and $10^4$ satellites selected randomly, ensuring that neither the central BCG mass nor its sky position were decisive factors. The objective is to evaluate the dependence of the measurements on the number of targets and the convolution effects on them. The measured satellite number density profile, $\Sigma_{\rm sat}$, for the four subsamples is shown in Fig.~\ref{fig.zou_vs_members} (left panel). In orange, the satellite density as a function of distance from the cluster centre is plotted for the four cases. In blue, the same satellite density profiles are displayed convolved with a Gaussian filter of $\sigma_{\rm pos}=2.4''$, as in the lensing measurements.

Even in the extreme case of just 100 targets, the outer region ($>10''$) is well recovered and there is little difference between the filtered and unfiltered cases, a difference that becomes negligible as the number of targets increases. Below $2$--$3''$ there is a pronounced inner break radius, clearly seen with the full sample. This absence of satellite galaxies is related to the finite size of galaxies, which causes them to overlap in projection, creating an unavoidable selection effect \citep{VAN05}. Interestingly, there are always a few pairs at very small angular separations, even for the smallest subsample. The convolution of these few central cases produces a distinctive bump in the central region, which can also arise without them if the inner break radius is comparable to or smaller than the convolution kernel. This central excess has been seen and discussed in previous works \citep{WAT10, WAT12, BUD12, TAL12, van2015evidence, van2016stellar}. Crucially, the convolved radial distribution remains continuous at all scales, showing only small fluctuations at intermediate angular separations for the smallest subsample due to the poorer statistics. This continuity, with no depression at scales corresponding to the \textit{Einstein Gap}, is the key result of this analysis: if a mass deficit were responsible for the lensing signal drop, it would necessarily leave an imprint here.

We therefore use $\Sigma_{\rm sat}$ as a smooth, continuous reference profile against which to compare the lensing measurements, and proceed to do so in the following paragraph. The results of this comparison are shown in the right-top panel of Fig.~\ref{fig.zou_vs_members}. To compare the mass density profile derived from the lensing results with the spatial distribution of the satellites, a normalisation must be performed, as both profiles are directly related. We note that our comparison between the magnification bias and satellite profiles is qualitative in nature and does not involve a joint physical fit with common free parameters. The satellite signal is rescaled via a single normalisation factor, estimated using a Markov Chain Monte Carlo (MCMC) approach, to match the amplitude of the magnification profile. This procedure is intended to assess whether both observables trace a similar underlying mass distribution, rather than to derive physical parameters from a joint model.

Once the satellite density profiles are renormalised using the outer region as a reference (between $10''$ and $100''$), we can compare the effect of using different numbers of satellites. Ignoring the smallest scales, the \textit{Einstein Gap} visible in the lensing results is not reflected in $\Sigma_{\rm sat}$ at any subsample size. Between 70 and 600~kpc both profiles follow a similar trend, especially when using the full satellite sample. The right-bottom panel of Fig.~\ref{fig.zou_vs_members} shows the difference between using the total number of satellites and each smaller subsample, supporting that robust results are achievable even with fewer satellites, particularly in the outer regions. Although oscillations in the signal emerge at scales similar to those of the \textit{Einstein Gap} for the smallest subsamples, they are never sufficiently strong nor sufficiently stable in position to constitute a plausible explanation for the lensing deficit. We therefore conclude that $\Sigma_{\rm sat}$ strongly disfavours both hypotheses proposed in CRE24: that the \textit{Einstein Gap} arises from a genuine lack of mass at those scales, or that it is a consequence of an insufficient number of lensing events. The signal deficit must instead have a lensing-specific origin, as explored in Section~\ref{sec:eins}.

\subsection{Dependence of the lensing signal and satellite distribution on BCG stellar mass}
\label{sec:massbins}

The comparison between $\Sigma_{\rm sat}$ and the lensing-derived mass density profile presented in the previous subsection was performed for the full cluster sample. However, the prominence of the \textit{Einstein Gap} is expected to depend primarily on the total halo mass and its central concentration, since these are the quantities that govern the strength of the lensing effect. Since the BCG stellar mass serves as an observational proxy for the total halo mass through the stellar-to-halo mass relation \citep[SHMR;][]{MOS13}, we repeat the analysis separating both the lensing measurements and the satellite density profiles into the seven BCG stellar mass bins defined in Table~\ref{tableMass}. This allows us to assess whether the signal deficit scales with the total halo mass, which would further support a lensing-specific origin for the \textit{Einstein Gap}, and to obtain improved CCF fits for each mass regime using $\Sigma_{\rm sat}$ as a low-noise proxy for the underlying matter distribution.

\begin{figure}
    \centering
    \includegraphics[width=0.87\textwidth]{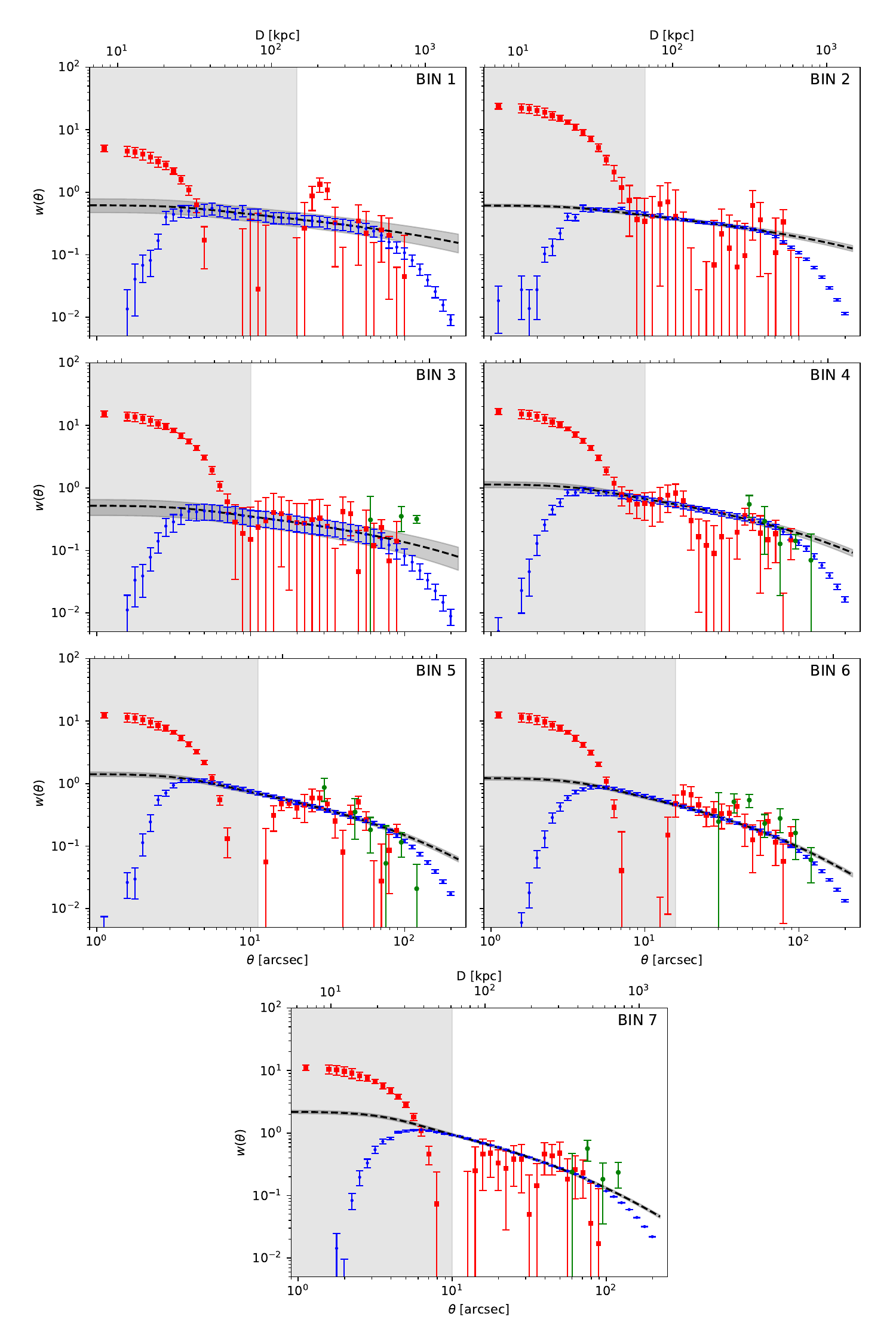}
\caption{Mass density profile analysis for each mass bin (PEAK centring). The panels are ordered by BCG stellar mass range from top to bottom and left to right, BIN1 being at the top left and BIN7 being the panel at the bottom. The red circles represent the results obtained using the lensing method, the green ones using the traditional estimator, while the blue squares correspond to the renormalized $\Sigma_{\rm sat}$ points. The best NFW fit to the outer region of the satellite number density profile for each BCG stellar mass range is represented by a dashed black line, and the shaded dark black region indicates its 68\% uncertainty. Measurements within the light gray areas were discarded for this analysis. }
    \label{fig.BINmasa_PEAK}
\end{figure}

\subsubsection{Profile comparison and NFW fits}
For this analysis, we follow the same procedure as in the overall case, but only for those galaxy clusters with a BCG with mass within the mass bin interval. Table \ref{tableMass} lists the number of lenses in each mass bin used for the analysis, as only those located within G09, G12, G15, and NGP fields are included. The lensing results of this analysis are shown in red in Fig. \ref{fig.BINmasa_PEAK}, where each panel denotes a different BCG stellar mass bin. Additionally, a satellite density stacking map can be created by separating satellites according to the same BCG stellar mass ranges. This will allow us to obtain the spatial distribution of the satellites and observe how it varies as a function of the central mass of the cluster.

Similar to the process described in Section \ref{sec:sat_ndp}, we can normalize $\Sigma_{\rm sat}$ based on the lensing data, which, due to the small number of lenses, exhibits large error bars.  The normalized satellite density profiles are shown in Fig. \ref{fig.BINmasa_PEAK} as blue points and they resemble the same trend as the lensing measurements providing evidence that both trace a similar mass density distribution. As they have much lower uncertainties and  show fewer oscillations than the data derived from lensing, these data can be utilised to refine the mass density profile fitting to a standard dark matter distribution, specifically the projected NFW\footnote{Further details are provided in Appendix \ref{sec:NFW}}.

For each BCG stellar mass interval, we determine the best normalisation factor, $p$, that rescales the satellite density profile to the lensing one, considering only those angular scales around $10''$ (measurements inside the gray areas in Fig. \ref{fig.BINmasa_PEAK} were discarded for this analysis). The fit is performed through an MCMC approach applied to the residuals between the lensing profile and the satellite profile. In this procedure, the covariance matrix of the residuals includes the individual covariance matrices of the lensing and satellite profiles, as well as their cross-covariance, which is estimated directly from the corresponding jackknife realisations of both profiles, using matched radial intervals and the same jackknife sectors.

Once the best-fitting value of $p$ is obtained, the satellite profile is renormalised, and its covariance matrix is consistently rescaled. The renormalised satellite profile is then used as a reference for the projected mass density profile to fit a projected NFW model and derive the best-fit mass and concentration parameters listed in Table \ref{tableMass}. The best NFW fit for each mass range is represented by a dashed black line in Fig. \ref{fig.BINmasa_PEAK}, while the shaded black region shows the corresponding 68\% credible interval. The NFW fit was also repeated using the upper and lower credible limits of $p$ in order to take into account the uncertainty in the normalization, and the resulting variation was included as an additional contribution to the uncertainties of the final parameter.

\subsubsection{Concentration values and the satellite contribution}
\label{sec:lowC}
Both magnification bias and the satellite distribution allow us to disentangle the contribution of satellites from that of the central region. As a result, the estimated halo masses and concentrations should be regarded as effective parameters that quantify the lensing strength associated with the satellite component, rather than direct measurements of the physical masses of the satellites.
This interpretation is also broadly consistent with galaxy-galaxy and cluster weak-lensing studies that model separately the central galaxy, satellite, and host-halo contributions \citep{LI09,HUD15,ZU15,UIT16,VEL17,LUO22}. These analyses generally find that the lensing contribution associated with satellite galaxies becomes most relevant at projected separations $>100-200$ kpc for BCG masses $M_\star < 10^{10.5-11.0}\,M_\odot$. The scales over which the satellite component is detected in our measurements are therefore comparable to those identified in independent shear analyses, despite the different observational techniques employed. Therefore, our estimates cannot be directly compared with standard concentration--mass ($C$--$M$) relations, which include the full halo contribution. As discussed in more detail in FER22, imposing a particular $C$--$M$ relation or forcing $C>2$ produces either an excess at the smallest angular separations or an underestimation at the largest ones.

The low concentration values are nonetheless fully consistent with the literature once this distinction is taken into account. Satellite galaxies are known to follow more extended profiles than the underlying dark matter, with typical concentrations of $C\sim2$--$3$ compared to $C\sim5$--$10$ for the DM halo \citep{NAG05, van2015evidence}. Our values are even lower, but this is expected: \cite{van2015evidence} showed that when the central excess is properly accounted for via a generalised NFW profile \citep{ZHAO1996, WYI01}, the resulting concentrations drop to $C\sim0.2$--$1.2$, in excellent agreement with ours. Similarly, \cite{LUO22} demonstrated that a higher overall satellite fraction adds an upward signal at larger radial scales, effectively "flattening" the total excess surface density profile. By contrast, a bias towards dynamically younger blue galaxies can be discounted, as red-sequence galaxies completely dominate the stellar mass distribution at the physical scales relevant to our analysis \citep{GUO13, HEN17}.

The results at the smallest angular separations reinforce our previous findings, further supporting the necessity of more than one mass density profile to fully describe the measurements at all angular separations (see FER22, CRE22 and CRE24 for a detailed discussion and different tests). This conclusion is consistent with a broader body of work highlighting the distinct lensing contributions of central and satellite galaxies. \cite{LI09} demonstrated that central and satellite galaxies produce very different lensing signals and must therefore be modelled separately, a point further developed by \cite{ZU15}, who showed that for stellar masses $M_\star < 10^{11}\,M_\odot$ the satellite contribution dominates the lensing signal at intermediate scales ($0.1$--$1$~Mpc). More directly relevant to our findings, \cite{SCH10b} showed that while the outer weak-lensing signal is well described by an NFW profile, the inner dynamical mass systematically exceeds the inward extrapolation of that same NFW fit even after subtracting the estimated stellar mass contribution, and that this excess cannot be accounted for by a model lacking a strong lensing component in the central region. Taken together, these results corroborate that the single-profile description breaks down in the inner region and that the central excess observed here reflects a genuine physical feature of the gravitational lensing rather than a methodological artefact.

Finally, as can be seen in Table \ref{tableMass} for both centring alternatives, the total halo masses derived from the NFW fit, $M_{NFW}$, deviate from the natural trend observed in other SHMR studies, $M_{SHMR}$ \citep[to cite some relevant ones,][]{MAN06,LEA12,MOS13,CON07,PIZ05}. The results for stellar masses below $10^{11.1}\,M_\odot$ (corresponding to BIN 1 through BIN 4) significantly exceed the expected SHMR. In contrast, for higher mass ranges (specifically BIN 5 and BIN 6), the results are consistent with the SHMR. The only exception is BIN 7, whose estimated halo mass falls below the SHMR, as expected if the measured signal is dominated by the satellite contribution. As already mentioned, these fitted masses act as effective parameters that quantify the lensing strength associated with the satellite contribution. Their diagnostic value lies precisely in the fact that, for several bins, they are substantially larger than expected from the SHMR. Such an excess indicates that the observed signal cannot be explained solely by the expected mass contribution of satellites to the cluster mass density profile and suggests the presence of additional effects, most notably miscentring and/or a stronger than expected lensing contribution from individual satellites \citep{MEN20,MEN22,MEN23}. These two non-exclusive possibilities are discussed in detail in the next subsection. 

\subsubsection{Miscentring and oscillatory behaviour}
\label{sec:miscenter}
Due to the discrepancies found in the mass results for BIN 1 through BIN 4, we aim to investigate in greater detail the potential explanations that could account for these findings. An analysis of the positions of the BCGs in lower-mass systems, specifically those with masses below $10^{11.1} M_\odot$, reveals a significant proportion of satellites with higher masses than the BCG. See Fig. \ref{fig.BIN1vsBIN7posv2} in Appendix \ref{ANX:figures} for a visual representation of satellite positions relative to the BCG centre, comparing the most extreme cases, Bin 1 and Bin 7. Moreover, Fig. \ref{fig.BINpercent}, also in Appendix \ref{ANX:figures}, shows the evolution of the percentage of satellites with masses exceeding that of the BCGs. This trend suggests that these are likely interacting galaxy groups, where the satellites are in equilibrium orbits prior to mergers occurring between the central galaxies. In these early cases, a significant proportion of satellites have greater masses than the BCG, which could produce a miscentring issue in the analysis. Therefore, it is normal that the galaxy satellites radial distributions are less concentrated in agreement with non-relaxed systems as the so called "non red sequence" galaxies \citep{GUO13,HEN17}.

In fact, in a normal situation, the SHMR indicates that the expected strength of the lensing effect of the outer part should be much weaker (lower $M_h$) than the observed one. Considering the high stellar masses of some of the satellites, one possible explanation for such high derived total masses could be the strong lensing effect of individual satellites in agreement with the recent findings of \cite{MEN20,MEN22,MEN23} that increase the lensing signal at intermediate scales. These studies found that cluster member galaxies produce galaxy-galaxy strong lensing events roughly an order of magnitude more often than predicted by cosmological simulations, implying that satellite substructures are more centrally compact than current models assume.

For clusters whose BCGs are in BIN4 and BIN5, the percentage of satellites with higher masses than the BCG is significantly lower. However, the satellite contribution to the halo mass is still above the SHMR indicating that their lensing contribution is still dominant or at least comparable to the BCG one. For the last bins of higher stellar mass, BIN6 and BIN7, we recover what should be the expected behaviour, i.e. that the satellites contribution is subdominant to the BCG one and our results are lower than the SHMR. We will return to this point for the BIN6 in the next section using the results from the magnification bias simulator.

The fact that both the lensing signal and the satellite distribution share the same trend at large scales indicates that the magnification bias signal is more closely related to the satellite population than to the smooth underlying DM halo in which they are embedded. This conclusion is further supported by the oscillatory behaviour of the lensing measurements, which is difficult to attribute to statistics alone. If the lensing signal at these scales is dominated by the cumulative effect of individual strong lensing events from relatively massive satellites, rather than the smooth weak lensing expected from the DM halo, significant oscillations can persist even after stacking. This is precisely because stacking tends to suppress only uncorrelated noise. Therefore, the low concentrations, the higher than SHMR masses in the lowest BCG stellar mass bins and the oscillatory features appear to share a common origin: a greater than expected strong lensing contribution from massive satellite galaxies within the foreground halos \citep{MEN20,MEN22,MEN23}. We stress that a dominant satellite contribution to the lensing signal at these intermediate separations does not imply a correspondingly large contribution to the mass density profile itself; rather, it reflects the cumulative effect of individual, strongly amplified lensing events, superimposed on a comparatively modest underlying satellite mass budget. This possibility opens an interesting complementary opportunity to add more observational information to this open issue.

Finally, we also observe that the outermost region of $\Sigma_{\rm sat}$ does not appear to follow the NFW trend \citep{TAL12}. This discrepancy is likely attributable to the limited completeness of satellite counts at large cluster radii, as satellites at scales larger than $\sim$1~Mpc ($\sim$$150''$) are not accounted for in the catalogue.

\subsubsection{Correlation between the Einstein Gap, the satellite lensing contribution and BCG stellar mass}
\label{sec:bcgmass_corr}

To assess systematically how the choice of cluster centre affects the detectability of the \textit{Einstein Gap}, we compare the mass-binned results obtained using the PEAK and the BCG positions as the stacking centre (Table~\ref{tableMass} and Figs.~\ref{fig.BINmasa_PEAK} and \ref{fig.BINmasa}). For BIN1 through BIN4, the two centrings give consistent results: neither shows a clearly defined gap, and both display comparable levels of oscillation in the lensing signal. This reflect the fact that a large fraction of clusters in this mass range are not fully relaxed (Section~\ref{sec:miscenter}), so neither centre can be regarded as an unbiased tracer of the underlying potential minimum. From BIN5 onward, however, the two centrings begin to diverge: the PEAK-centred stacks consistently produce a deeper and more sharply delimited \textit{Einstein Gap} than the BCG-centred ones. The most extreme case is BIN7, where the gap is essentially washed out under BCG centring despite the large halo mass involved (see Fig.~\ref{fig.BINmasa} in Appendix~\ref{ANX:BCG_centre}), while it remains clearly visible under PEAK centring. For this reason, we adopt PEAK centring as our fiducial choice throughout the remainder of this section and the following analysis, and 
we return to this specific case in Section~\ref{sec:str_len}.

The key empirical result of the mass-binned analysis is the fact that the depth and prominence of the \textit{Einstein Gap} increase systematically with BCG stellar mass using the PEAK centring (and therefore, via the SHMR, with total halo mass) as can be seen in Fig.~\ref{fig.BINmasa_PEAK}. In BIN1 the measurements show strong oscillation mostly at all the angular scales due to miscentring and the stochastic satellite contribution as already discussed, making difficult any meaningful conclusion about the presence of the \textit{Einstein Gap}. In BIN2 to BIN4, the results are more stable and there is no clear lack of signal around the $10''$. In fact, considering the total masses estimated for each mass bin assuming a traditional SHMR \citep[][see Table \ref{tableMass}]{MOS13}, the averaged expected Einstein Ring radius is below $2''$ and therefore not detectable with the current resolution.
From BIN5 the \textit{Einstein Gap} is clearly formed, being widest in BIN6. 
In all three bins the gap is highly significant, with $N_\sigma>29$. This significance is obtained by comparing the measured profile against a smooth reference model fitted to the angular bins outside the gap and propagating the full bin-to-bin Jackknife covariance\footnote{The significance of the gap is estimated by computing $\chi^2 = (\vec{d}-\vec{m})^{T} C^{-1} (\vec{d}-\vec{m})$ between the measured profile $\vec{d}$ (including the gap-affected angular bins) and a smooth reference model $\vec{m}$, obtained by fitting only the bins outside the gap region, where $C$ is the full bin-to-bin Jackknife covariance matrix (Section~\ref{sec:massbins}). This $\chi^2$ is converted into a $p$-value using the $\chi^2$ cumulative distribution with the appropriate effective number of degrees of freedom, which is then translated into a one-sided, Gaussian-equivalent significance, $N_\sigma$.}.
A systematic artifact (whether from the stacking procedure, the cross-matching strategy, or the positional uncertainty) would affect all mass bins equally and could not produce a mass-dependent signal. The scaling therefore constitutes strong evidence that the gap is physically real.

Moreover, the satellite profiles are smooth and continuous across all mass bins, with no depression at the scales corresponding to the gap. This strongly disfavours the interpretation that the gap reflects an actual lack of mass, indicating that this conclusion holds independently of the mass bin considered.

In addition, the oscillatory behaviour at intermediate scales is most prominent in low-mass systems, where satellites represent a larger fraction of the total lensing signal, and become progressively less evident towards the highest BCG stellar mass bins, where the BCG itself dominates the central mass budget and the satellite contribution is subdominant, as can be seen in the smoothest profiles of the highest mass bins in Fig.~\ref{fig.BINmasa_PEAK}. This conclusion indicates that the \textit{Einstein Gap} cannot be considered a consequence of the stochastic contribution of the satellites.

Overall, since the \textit{Einstein Gap} scales with halo mass and cannot be explained by either methodology, a mass deficit or satellite stochastic contribution, its origin must be a lensing-specific physical mechanism, most probably strong lensing considering the angular separations and lens masses under consideration. Identifying that mechanism is the goal of the next section.

\section{The \textit{Einstein Gap}: a strong lensing feature of massive halos}
\label{sec:eins}

Several of our previous studies report an unexpected drop in lensing signal at intermediate angular scales, typically around 10--$30''$, which coincides with the lower limit used in weak lensing analyses based on the cross-correlation function. This anomaly appears consistently across various lens samples, including GAMA, SDSS, and QSO catalogues and methodologies -- galaxy-galaxy lensing \citep{HUD98, HOE03, HOE05, MAN06,Man09,SHE09, LI09, ZU15,UIT16,VEL17,LUO17,LUO22, MIS23,XU24}, weak lensing from the WHL12 cluster catalogue \citep{BAU14}, shear profile stacking \citep{JOH07, HUD15, RAN22}, and joint analyses combining strong+weak lensing \citep[e.g.][]{UME16, NEW13}. This scale also corresponds to the transition region between the dark matter halos of clusters and those of the BCGs, where previous works \citep[e.g.][]{GAV07} observed a similar dip. A comparable signal drop was noted by \cite{MEN10}, though attributed to dust extinction, an explanation that does not apply to our infrared data, where attenuation is negligible. 
Crucially, in all of the above cases a comparable feature appears to be present in the published figures of several of these studies, although none of them pursued it further or offered a physical interpretation.
The \textit{Einstein Gap} is therefore not a novelty of our pipeline but a recurring, unexplained feature of gravitational lensing mass profiles across a wide range of lens types, redshifts, and methodologies.

The observational tests presented in Section~\ref{sec:method} collectively support the conclusion that the \textit{Einstein Gap} cannot be attributed to systematics, methodological choices, or data artefacts, and that its prominence scales systematically with BCG stellar mass, and therefore with total halo mass via the SHMR. These properties collectively point toward a lensing-specific physical origin, most probably associated with strong lensing effects given the angular separations and halo masses under consideration. Before demonstrating this through an improved magnification bias simulator that fully incorporates gravitational lensing via the lens equation, we first summarise in Section~\ref{sec:ruling_out} the observational evidence that strongly disfavours the remaining alternative explanations.

\subsection{Disfavouring alternative explanations}
\label{sec:ruling_out}
\subsubsection{Why alternative observational systematics cannot generate the Einstein Gap}
An initial concern is whether the deficit could instead originate from the background SMG sample itself, or from the H-ATLAS/WISE detection process, rather than being a genuine property of the foreground mass distribution. This point was already discussed in previous works \citep{FER22,CRE24,GON26}, but we summarise here the main conclusions.

\textit{Magnification-dependent selection and completeness.} A deficit could in principle arise if the flux-limited detection of background sources responded non-linearly to magnification in a way that mimics a gap. This possibility is directly tested by the improved simulator (see Section~\ref{sec:simulator}), since the same synthetic background-source pipeline (number counts from \cite{CAI13}, the SPIRE $250\,\mu$m beam, and the H-ATLAS flux limit) underlies both (i) the weak-lensing-only simulator of CRE22 and CRE24 and (ii) the present ray-tracing version; the two differ only in whether the lens equation is solved for the source-plane displacement. The weak-lensing-only simulator, despite sharing an identical detection pipeline and selection function, never produced a gap-like feature at any lens mass; the deficit appears only once the displacement of images near the Einstein ring is included, and only for sufficiently massive lenses (Section~\ref{sec:simulator}). This offers a direct comparative evaluation of the detection pipeline in isolation from the lensing model, and it shows that non-linear selection or flux-boosting effects, on their own, cannot reproduce the observed deficit. Moreover, under the weak lensing approximation magnification varies smoothly and monotonically with angular separation, so there is no transition scale at which a magnification-driven selection effect could produce a localised deficit; the only genuine transition scale in the problem is the caustic structure associated with the Einstein ring itself (Section~\ref{sec:SISSA}), which is itself a strong lensing feature rather than an alternative to it.

\textit{Blending, beam confusion, and catalogue-intrinsic bias.} A related concern is whether beam confusion or positional systematics near massive foreground systems could suppress the background source counts at the relevant scales, particularly given that the gap's angular extent is comparable to the SPIRE $250\,\mu$m beam (${\rm FWHM}=17.8''$, \cite{VAL16, BOU16,MAD18}). Several independent lines of evidence argue against this. First, our lens and source populations are selected to have essentially null cross-contamination: the foreground clusters and their BCGs are optically bright but negligible far-IR emitters, while the background SMGs are selected precisely for being bright in the far-IR and comparatively faint in the optical. This decoupling is what allows the cross-correlation to probe angular scales well inside the physical extent of the lens light without the blending and shredding issues that affect galaxy-galaxy lensing studies based on a single optical band. Second, the beam itself is fully propagated through the simulator (Section~\ref{sec:simulator}), where it acts only as a smoothing term on the recovered profile, as already established in CRE24, and does not by itself generate a gap in the weak-lensing-only case discussed above. Third, and most decisively, improving the source positions via the WISE cross-match \emph{strengthens} rather than erases the gap (Fig.~\ref{fig:methods}); a beam-confusion or blending artefact tied to the native Herschel resolution should instead be washed out, not enhanced, once the effective positional accuracy is improved by an order of magnitude. The same conclusion follows from Appendix~\ref{ANX:smooth}, where the gap becomes more prominent as the smoothing kernel is reduced, and from Appendix~\ref{ANX:xmatch}, where three different WISE cross-matching criteria (nearest, farthest, random counterpart) all recover the same feature, ruling out a bias tied to a specific counterpart-selection choice. Finally, if the deficit were intrinsic to the background catalogue or its cross-match with WISE, it should appear with the same strength regardless of the properties of the foreground lens population, since the background sample and detection pipeline are identical in every bin. This is directly contradicted by the mass-binned analysis (Fig.~\ref{fig.BINmasa_PEAK}) and by the simulator comparison (Fig.~\ref{fig.simulator}): the depth and angular position of the gap scale systematically with BCG stellar mass and, correspondingly, with the halo mass required to reproduce it, exactly as expected if the effect originates in the lens rather than in the source catalogue.

Taken together, these tests show that the \textit{Einstein Gap} cannot be attributed to an incompleteness depending directly on magnification, beam confusion, cross-matching bias, or any systematic intrinsic to the SMG background sample. We note that the feature is not unique to this background sample or to this observational technique: a comparable signal deficit at similar angular and physical scales has been noted, without further investigation, in galaxy-galaxy and cluster weak lensing studies that employ an entirely different lensing observable (shear rather than magnification bias) and an entirely different background population (optically selected galaxies rather than far-IR SMGs; see Section~\ref{sec:intro} and the opening of Section~\ref{sec:eins}, and references therein). Moreover, the observed mass-scaling is a discriminating signature of a genuine, mass-dependent, lens-side effect.
Therefore, a systematic specific to the H-ATLAS/WISE catalogue or to our cross-correlation pipeline cannot account for a feature that recurs across independent instruments, bands, background populations, and lensing observables. A shared physical origin in the foreground lens population is by far the more economical explanation.

\subsubsection{Assessing the alternatives identified in CRE24}
Three candidate explanations for the \textit{Einstein Gap} were identified in CRE24 and have been systematically tested through the observational analysis presented in Section~\ref{sec:method}. We summarise the evidence against each of them here.

\textit{Statistical artifact.} A preliminary explanation was that the number of lensing objects might be insufficient for a robust statistical analysis, producing a spurious signal deficit. Two lines of evidence argue against this. First, if the deficit were purely statistical, its magnitude should scale directly with the number of lenses, since the number of potential background sources is the same for all bins. However, as shown in Fig.~\ref{fig.BINmasa_PEAK}, the depth of the gap correlates far more strongly with the BCG stellar mass, and therefore with total halo mass via the SHMR, than with the sample size. This is supported by our quantitative analysis in Section~\ref{sec:bcgmass_corr}, where we find the gap detected at $N_\sigma>29$ across BIN5--BIN7, using the full Jackknife covariance matrix of the stacked profile. 
This is particularly evident in the comparison between BIN4 and BIN5: BIN4 contains roughly half as many lenses as BIN5, yet exhibits only a negligible signal deficit around $10''$. An independent argument comes from the smoothing-kernel analysis presented in Appendix~\ref{ANX:smooth}, the gap becomes progressively more prominent and better defined as the positional smoothing kernel is reduced. A statistical fluctuation would not exhibit this behaviour: reducing the smoothing increases the noise level, which would cause any spurious feature to become less significant rather than more coherent. The fact that the opposite is observed is therefore a clear signature of a genuine physical feature in the data. Together, these two results strongly disfavour a statistical origin for the \textit{Einstein Gap}.

\textit{Methodological artifact.} To ensure that the signal drop is not an artifact of the stacking method or of the specific cross-matching strategy, we repeated the analysis using three independent estimators of the CCF (the traditional Landy-Szalay estimator, stacking in angular distance, and stacking in physical distance) as described in Section~\ref{sec:stack} and shown in Fig.~\ref{fig:methods}. Additionally, we tested three different cross-matching criteria between the H-ATLAS and WISE catalogues (nearest, farthest, and random counterpart within $17.6''$), as detailed in Appendix~\ref{ANX:xmatch}. The \textit{Einstein Gap} appears consistently across all configurations, strongly disfavouring any methodological origin.

\textit{Genuine absence of mass.} Perhaps the most physically meaningful alternative is that the gap reflects a true deficit of matter at intermediate scales ($\sim$50~kpc). If so, a corresponding depression would necessarily appear in $\Sigma_{\rm sat}$, which serves as a direct, lensing-independent tracer of the underlying mass distribution. As demonstrated in Section~\ref{sec:sat_ndp} and Fig.~\ref{fig.zou_vs_members}, $\Sigma_{\rm sat}$ is smooth and continuous at all scales, with no depression at the scales corresponding to the gap, regardless of subsample size or BCG stellar mass bin. This strongly disfavours a genuine mass deficit as the cause of the observed lensing signal drop.

Although none of the three tests above is individually decisive, their combined results, together with the positive reproduction of the gap presented below, strongly favour a lensing-specific origin for the \textit{Einstein Gap}. The most plausible explanation is the onset of strong lensing effects, specifically, the displacement of background source positions in the strong lensing regime, an effect not fully incorporated in the CRE24 simulations. We test this hypothesis by enhancing the simulator to solve the full lens equation, as described in the following subsection.

\subsection{A magnification bias simulator with full ray-tracing}
\label{sec:simulator}

To enable a more reliable analysis of the results and better understand the data behaviour, we developed a magnification bias simulator. In its preliminary version, the simulator modelled foreground lenses using an NFW mass density profile (see Appendix~\ref{sec:NFW}) and estimated how these lenses magnify the emission of background sources at a given redshift. This initial approach focused only on total magnification, without accounting for other lensing effects in the strong lensing regime, such as the apparent displacement of background source positions. The first results from this simulator were presented in CRE24, assuming an average lens mass of $10^{14}\,M_\odot$ and a lens redshift of $z=0.3$. These results revealed discrepancies with the observed data: the expected signal deficit was absent, and instead a deficit appeared at smaller scales corresponding to the central galaxy region.

\begin{figure}
    \centering
    \includegraphics[width=0.49\columnwidth]{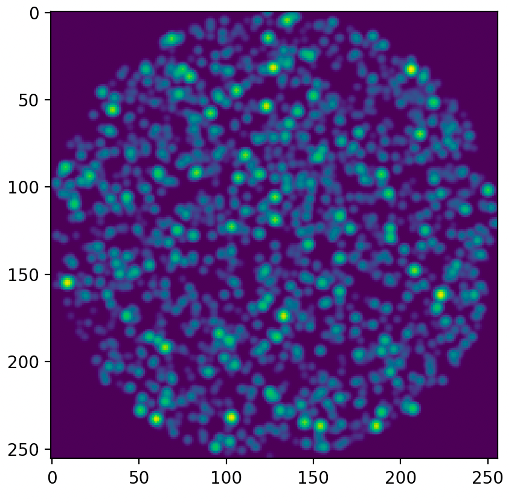}
    \includegraphics[width=0.49\columnwidth]{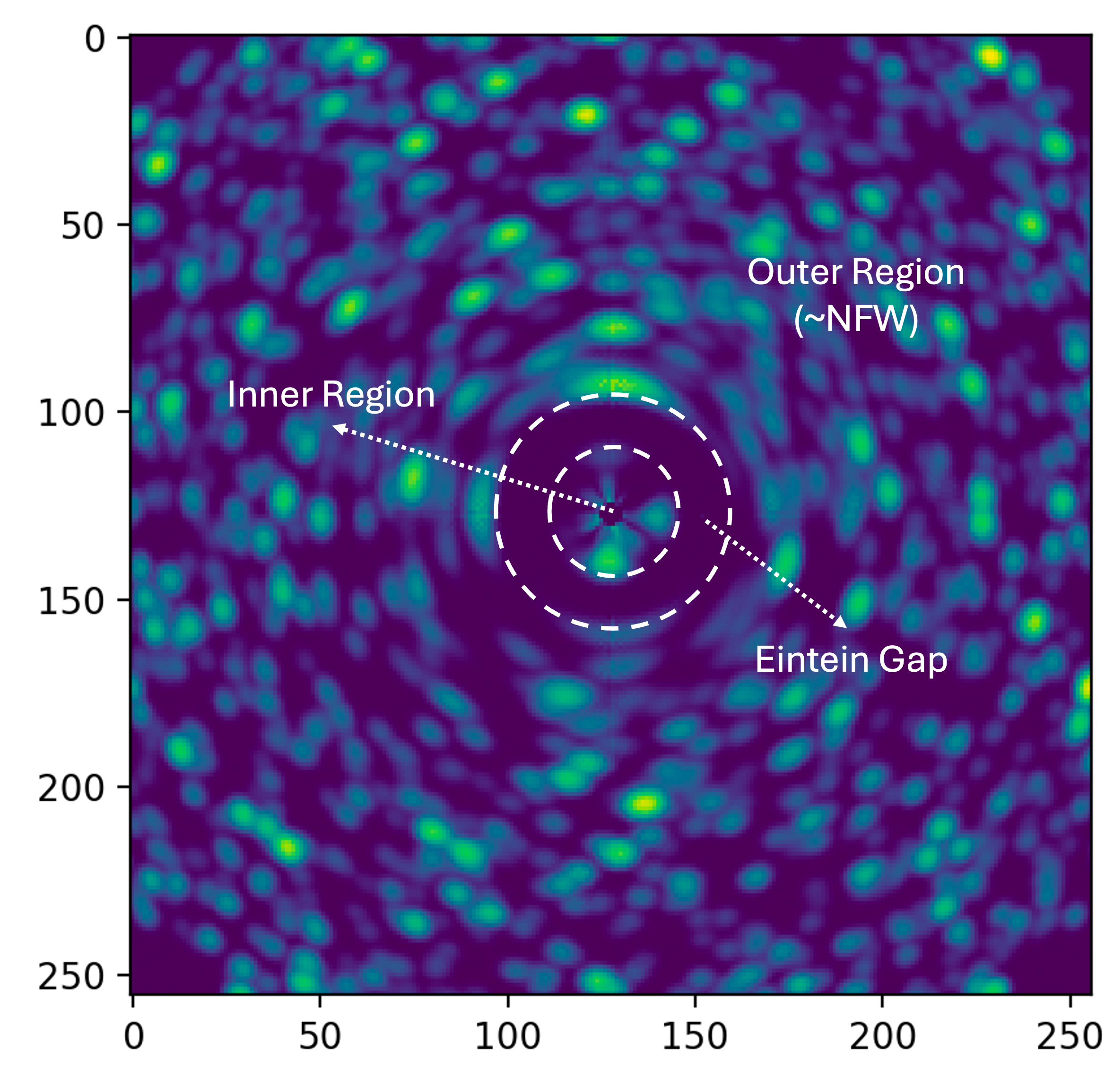}
    
    \caption{Realistic simulation of a background sample visually showing the strong lensing effect produced by a central object with $M_h = 7 \cdot 10^{13} M_\odot$ and a SISSA mass density profile (see text for more details). The left panel shows the configuration without the lens, while the right panel includes the lens, highlighting the resulting strong lensing effects and, in particular, the \textit{Einstein Gap} just inside the Einstein ring.}
    \label{fig.stakinglens}
\end{figure}

To address these discrepancies, we introduced two improvements. First, we adopted a more sophisticated mass density profile, implementing the SISSA profile \citep{LAP12}, which provides a simple parametric form combining an NFW profile for the dark matter component with a Sérsic profile for the stellar component; further details are given in Appendix~\ref{sec:SISSA}. Second, we enhanced the simulator by fully incorporating gravitational lensing effects via the lens equation, which alters the positions, shapes, and brightness of background images, going beyond the purely magnification-based approach of CRE24.

The effect of these improvements is demonstrated in Fig.~\ref{fig.stakinglens}, which shows stacking maps with and without a lens of given mass.\footnote{This figure is for illustrative purposes only. No detection threshold is applied and the colour scale is logarithmic in order to reveal the full extent of the strong lensing effects.} In particular, the strong lensing regime near the centre of the lens creates a characteristic empty annulus, the \textit{Einstein Gap}, bounded by the Einstein ring on the outside and by the inner critical curve, where secondary images form, on the inside \citep{LAP12}. The \textit{Einstein Gap} is therefore a direct consequence of strong lensing even for a single isolated lens.

In practice, however, detecting this feature requires stacking a large number of lenses. With the current datasets, only a few background sources are expected within the relevant angular area for any individual lens, and the complexity of the mass distribution in galaxy clusters further complicates the identification of such a feature in individual systems. To render the \textit{Einstein Gap} statistically detectable, hundreds of lenses with similar physical properties (in particular, comparable mass and redshift distributions) must be stacked, so that the signal coherently accumulates without being blurred by the diversity of the lens population.

Background sources were simulated within a $100''$ radius (stacking maps of 400 pixels with a pixel size of $0.5''$) using the \texttt{CORRSKY} software\footnote{Zenodo DOI: https://doi.org/10.5281/zenodo.20201113} \citep{GN05}, incorporating source number counts from \cite{CAI13} and the angular power spectrum from \cite{LAP11}. For the lenses, we adopted parameters from Table~1 of \cite{LAP12} to describe the SISSA profile, assuming $\log\Sigma_0 = 9.5$ and $\eta = 0.8$, corresponding to a dark-matter-to-stellar mass ratio of $\sim$30, a Sérsic index of $\sim$4, and a halo concentration of $\sim$5. As noted in that work, both $\Sigma_0$ and $\eta$ exhibit only weak dependence on the mass distribution parameters for a fixed total halo mass (see Appendix~\ref{sec:SISSA}), so the total halo mass remained the sole free parameter in this study. We assumed the ZOU redshift distribution, incorporating the resulting angular distance variations, and considered all lenses to share the same total mass\footnote{The current simulator assumes a single halo mass per stack because it is designed to establish the existence and mass-dependence of the gap mechanism, not to reproduce the stacking-induced broadening from a realistic mass distribution. The potential redshift broadening, by contrast, is already taken into account, since the simulator incorporates the ZOU redshift distribution and its associated angular-diameter-distance variations.}.

For each simulation, we computed the lensing effect of the lens on the source map, convolved the lensed map with the SPIRE $250\,\mu$m beam \citep[${\rm FWHM} = 17.8''$,][]{VAL16}, extracted the positions of sources brighter than the H-ATLAS detection limit of $\sim$28~mJy \citep{VAL16, MAD18}, and recorded those positions in the corresponding pixel of the stacked map. This sequence replicates the Herschel catalogue production steps applied to our background sample. The resulting simulated stacked maps were then smoothed with a $2.4''$ Gaussian kernel to replicate the positional uncertainty of the CCF measurements, and analysed following the same methodology as the real data.

The simulations show that for lens masses $M_{\rm H} \lesssim 4\times10^{13} M_\odot$ the results are consistent with those of the previous simulator, with negligible strong lensing effects. Above this threshold, increasing the total lens mass leads to the emergence of a ring-shaped signal deficit in the stacking maps, the \textit{Einstein Gap}, whose angular position scales with halo mass as $\theta_E \propto M_{\rm H}^{2/3}$ \citep{LAP12} as for the Einstein ring. This mass-dependent scaling provides a natural explanation for the correlation between the prominence of the \textit{Einstein Gap} and the BCG stellar mass discussed in Section~\ref{sec:bcgmass_corr}: more massive halos produce a larger Einstein radius and therefore a more prominent and better-defined gap.

Finally, we used the simulator to validate the methodology employed in Section~\ref{sec:sat_ndp} to derive the outer-region halo mass from the normalised satellite density profile and NFW fit. Since the current simulator considers only a single central lens with no satellite contribution, at halo masses $M_{\rm H} \gtrsim 5\times10^{13}\,M_\odot$ the simulated profile is steeper than the low-concentration profiles observed in the satellite distributions (Fig.~\ref{fig.BINmasa_PEAK}), and the methodology recovers only a fraction of the total mass, as expected. At $M_{\rm H} \sim 10^{13}\,M_\odot$, where the profile closely follows an NFW form, the input mass is recovered accurately. At even lower masses an upward small sample bias appears: for example, a simulated lens with $M_{\rm H} = 10^{12}\,M_\odot$ yields a recovered mass of $\sim$$4\times10^{12}\,M_\odot$. This indicates that the methodology does not introduce spurious high-mass estimates at the scales relevant to our analysis, and provides an additional line of evidence that the inflated halo masses derived for BIN1--BIN4 arise from the additional lensing contribution of massive satellites rather than from a methodological bias, consistently producing lower concentration values as discussed in Section~\ref{sec:lowC}.

\begin{figure}
    \centering
    \includegraphics[width=\textwidth]{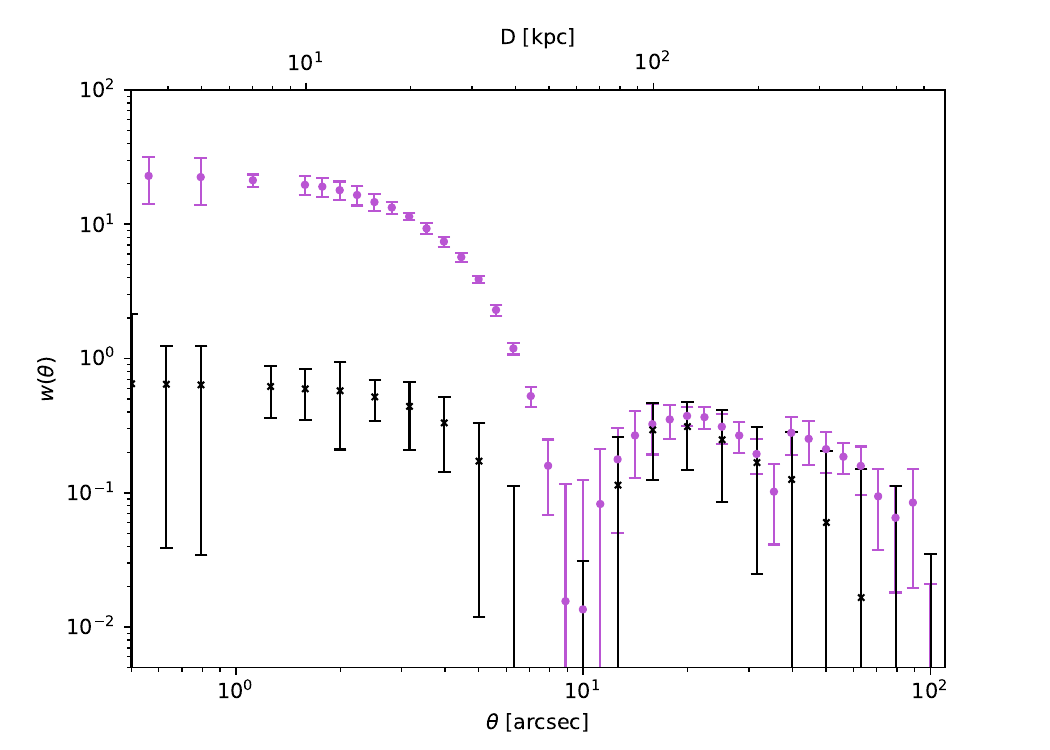}
    \caption{Comparison between simulated mass density profile and the measured one from the whole ZOU clusters catalogue. The black crosses represent simulation results using a halo mass of $\sim 5 \times 10^{14} M_\odot$, while violet circles correspond to observational data. Comparing the results for the whole sample with those of the individual mass bins (Fig. \ref{fig.BINmasa_PEAK}) provides evidence for the erosion/dilution mechanism affecting the \textit{Einstein Gap} in the stacking analysis (see Section \ref{sec:str_len} or a more detailed discussion).}
    \label{fig.simulator}
\end{figure}

\subsection{Strong lensing origin of the \textit{Einstein Gap}}
\label{sec:str_len}
When comparing the simulator results with the observations using the ZOU catalogue as the lens sample (the gap is detected with a significance of $N_\sigma= 12.6$; see Fig.~\ref{fig.simulator}), we find good agreement between the external lensing measurements ($>10''$) and the angular position of the \textit{Einstein Gap} for an average cluster halo mass of approximately $M_{\rm H}\sim5\times10^{14}\,M_\odot$. We stress that the case for a strong-lensing origin of the \textit{Einstein Gap} does not rest on this specific mass value; it follows instead from the qualitative and structural tests presented above: the gap appears only once the lens equation is solved and disappears in an otherwise identical weak-lensing-only simulator (Section \ref{sec:ruling_out}), and its position and depth scale with BCG stellar mass in the manner predicted by $\theta_E \propto M_{\rm H}^{2/3}$ (Section \ref{sec:bcgmass_corr}), a differential test of the mechanism that does not rely on the estimated absolute mass value being numerically exact.
The recovered halo mass should instead be regarded as a first illustration that the \textit{Einstein Gap} can, in principle, be used as a diagnostic of the total lens mass, once the parameters of the mass density profile (concentration, dark-matter-to-stellar mass ratio, satellite contribution) are properly modelled. However, deriving a precise mass estimate is beyond the scope of the present simplified simulator. 

As a simple sanity check, we note that BIN6, the most representative bin of the whole ZOU sample with $\log(M_\star/M_\odot)\sim11.53$, has an SHMR-expected halo mass of $\log(M_h/M_\odot)\sim{14.83}$ (Table~\ref{tableMass}), consistent with the value recovered from the simulator. Moreover, the resulting Einstein radius, $15$--$25''$ for a SISSA profile with the adopted parameters \citep{LAP12}, corresponds closely to the outer boundary of the observed gap, illustrating that the recovered mass is not merely a free normalisation but yields Einstein-radius values of the expected order for a massive cluster halo. We regard this agreement as encouraging but secondary, given the physical simplifications of the current implementation.

At first sight, the persistence of a relatively well-defined gap after stacking lenses spanning a broad range of masses and redshifts may seem surprising, since the corresponding Einstein radii should vary substantially from object to object, and one might expect this diversity to smear the feature out. This expectation, however, overlooks both the intrinsically sparse nature of the magnification bias signal and its strong dependence on lens mass. The probability of a lens-source association in our sample is very low, approximately one pair per $10^3$ lenses \citep{LAP12}, so the measured correlation at any given separation is produced by only a small subset of successful strong-lensing events. Moreover, because the pair rate grows steeply with lens mass through the magnification-dependent flux boosting required to place a source above the H-ATLAS detection limit, the stacked signal is preferentially weighted toward the most massive lenses near the peak of the redshift distribution, rather than representing a simple average over the full population. These systems generate both the largest source displacements and the most pronounced deficits, while lower-mass lenses or those far from the redshift distribution peak contribute much more weakly and are therefore inefficient at refilling the regions depleted by the strongest lenses. Consequently, the deficit produced by the most efficient lenses is not easily compensated by the much weaker contribution of the majority of systems, allowing a well-defined gap to survive the stacking process despite the broad underlying distribution of lens masses, redshifts, and Einstein radii.

This same competition between the dominant, high-efficiency sub-population and the more numerous but weaker-lensing remainder also explains the observed width of the gap and its dependence on the mass-redshift range being stacked. Within a single BCG stellar mass bin ($\sim 0.3$ dex in $log(M_{BCG}/M_\odot)$; Table \ref{tableMass}), the internal mass dispersion would, via $\theta_E \propto M_{\rm H}^{2/3}$ \citep{LAP12}, naively predict a spread of order $\sim$1 dex in halo mass and hence a factor of $4-5$ in $\theta_E$ if all lenses in the bin contributed equally. The observed gap widths are instead narrower -- a factor of $\sim 3.5$ in BIN5 ($20$--$70$~kpc), $\sim 2.0$ in BIN6 ($40$--$80$~kpc), $\sim 1.8$ in BIN7 ($50$--$90$~kpc) -- indicating that the lower-mass, off-peak tail within each bin partially refills the boundaries of the deficit produced by the dominant sub-population, narrowing the observed feature relative to the simple full-bin expectation. This erosion effect is dramatically demonstrated when the mass range being stacked is widened further: the gap recovered using the entire ZOU sample (Fig. \ref{fig.simulator}), spanning the full $\sim 2$ dex range of BCG stellar mass, is narrower and less pronounced than any of the individual mass-bin detections, and is well described by a single effective halo mass ($\sim 5 \times 10^{14} M_\odot$) comparable to the highest-mass bins (BIN5–BIN6) alone. This is precisely the behaviour expected if the numerous low-mass systems (BIN1–BIN4), whose Einstein radii lie below our angular resolution and which therefore do not produce a resolvable gap on their own, contribute a smooth, gap-free signal that dilutes and narrows the feature produced by the higher-mass tail, without erasing it entirely thanks to the mass-dependent efficiency weighting described above.

The strong lensing interpretation also provides a natural explanation for the mass-dependent scaling of the \textit{Einstein Gap} identified in Section~\ref{sec:bcgmass_corr}. Since $\theta_E \propto M_{\rm H}^{2/3}$ \citep{LAP12}, the Einstein radius grows with halo mass, and the gap only becomes detectable once $\theta_E$ exceeds the positional smoothing scale of our measurements. For a typical SISSA profile at $z_l \sim 0.5$ and $z_s \sim 2.5$, a halo mass of $M_{\rm H} \sim10^{13}\,M_\odot$ yields $\theta_E \sim 2''$ \citep{LAP12}, well below our smoothing kernel, which explain again why the \textit{Einstein Gap} is not detectable in BIN1--BIN4. For higher masses, BIN5--BIN7, the Einstein radius grows sufficiently large to produce an observable gap, as seen in Fig.~\ref{fig.BINmasa_PEAK}. In these bins, both the inner and outer boundaries of the gap increase with mass (BIN5: $20$--$70$~kpc; BIN6: $40$--$80$~kpc; BIN7: $50$--$90$~kpc), as expected from the $\theta_E$--$M_{\rm H}$ scaling. The increase is more pronounced in the lower boundary, which we attribute to the growing satellite contribution at angular separations $\gtrsim 10''$ (or equivalently $\gtrsim 80$--$100$~kpc): at these scales the satellite lensing signal, which is not suppressed by the \textit{Einstein Gap} mechanism, begins to dominate the total profile \citep{LI09,HUD15,ZU15,UIT16,VEL17,LUO22}, effectively filling in the outer edge of the gap.

The case of BIN7 when using the BCG as the central position (see Fig.~\ref{fig.BINmasa} and Appendix~\ref{ANX:BCG_centre}) deserves separate discussion, as the \textit{Einstein Gap} is not clearly detected there despite the high halo masses involved. As discussed in Section~\ref{sec:miscenter} and Appendix~\ref{ANX:figures}, BCGs in this bin are more centrally located within their clusters and possess sufficiently high stellar masses to produce strong lensing effects on their own. Furthermore, these clusters are also the richest in our sample: the mean number of satellites per BCG increases monotonically across the mass bins (Table~\ref{tableMass}), from $\sim16$ in BIN1 to $\sim33$ in BIN7, roughly twice the richness of the lowest-mass bin. As already noted for BIN5--BIN7 above, the growing satellite population at these masses contributes a lensing signal at angular separations $\gtrsim10''$ that is not suppressed by the \textit{Einstein Gap} mechanism \citep{LI09, ZU15}, progressively filling in the outer edge of the gap as richness increases. In BIN7, this effect is strong enough, combined with the centrally concentrated BCG mass distribution, to produce a smoother magnification profile that suppresses the formation of a distinct \textit{Einstein Gap}. This is demonstrated by comparing the BCG and PEAK centring results: as shown in Fig.~\ref{fig.BINmasa_PEAK}, the PEAK-centred analysis yields a cleaner and more regular \textit{Einstein Gap} whose prominence increases more uniformly with stellar mass. This comparison demonstrates that the satellite contribution cannot be neglected when analysing the strong lensing regime, and that the choice of centre can significantly affect the detectability of the feature due to the internal structure and mass composition of the lensing systems \citep{MOL02}.

These results carry an important practical implication: traditional weak lensing analysis methodologies, which assume that the signal at all angular scales can be described by a smooth mass density profile, should be applied with caution in the inner region ($\lesssim 10''$) where strong lensing effects dominate. The findings of CRE24, for instance, while consistent with previously reported results in the literature, should be reinterpreted in this light, as the inner profiles presented there do not fully account for the strong lensing regime. More broadly, the inability of any single smooth profile to reproduce the inner extended emission is not a failure of the mass model but a direct consequence of the lensing potential near the inner caustic, which cannot be captured by a weak lensing approximation regardless of the profile assumed.

Discrepancies between the simulator and the observations nevertheless remain in the inner region, where the predicted signal falls below the observed one. This mismatch is most likely a consequence of two limitations of the current implementation. First, the rigidity of the SISSA power-law parametrization in the central region may underestimate the true concentration, and allowing a higher central concentration would probably bring the inner signal into better agreement. Second, the current simulator considers only a single central lens with no satellite contribution; including satellites with varying mass fractions, density profiles, and spatial distributions would modify the lensing potential and its caustics, affecting both the inner and outer signals \citep{MOL02}. A systematic exploration trough simulations of this extended parameter space would be computationally intensive and is beyond the scope of the present work, but motivates future studies with higher-fidelity lensing simulations that incorporate the full complexity of the cluster mass distribution.

Finally, the comparison between PEAK- and BCG-centred results opens a promising avenue for future investigations. By quantifying how the \textit{Einstein Gap} properties vary with the choice of centre, it may be possible to determine whether the lensing signal within groups and clusters originates from a common smooth halo or is instead dominated by the contributions of individual member galaxies \citep{MEN20, MEN22, MEN23}. Such an analysis would provide a new observational handle on the degree of substructure lensing in massive systems, complementing existing approaches based on arc statistics and flux ratio anomalies.


\section{Conclusions}
\label{sec:concl}
We investigate the origin of the \textit{Einstein Gap}, a characteristic signal deficit observed at intermediate angular scales ($\sim10''$, corresponding to $\sim50$ kpc) in magnification-bias lensing profiles. Although this feature was previously identified across all lens samples analysed in CRE24 and has appeared in numerous independent weak-lensing studies \citep{MAN06, JOH07, GAV07, MEN10, BAU14, UME16, XU24}, its physical origin remained unexplained. In this work, we resolve this question by ruling out statistical, methodological, and mass-deficit explanations, and by showing, through full ray-tracing simulations and its scaling with BCG stellar mass, that the \textit{Einstein Gap} is a genuine strong-lensing feature.

To identify the physical origin of the \textit{Einstein Gap}, we first argue against its potential origin as observational systematics or data artefacts, including detection-pipeline systematics specific to the H-ATLAS catalogue.
Then, we pursued a three-pronged observational strategy using galaxy clusters from the ZOU catalogue as lenses and high-redshift SMGs from H-ATLAS/WISE as background sources. First, we demonstrated the robustness of the feature across three independent estimators of the angular cross-correlation function and three different cross-matching strategies, indicating that it cannot be attributed to any methodological artifact. Second, we showed that the gap becomes progressively more prominent as the positional smoothing is reduced, disfavouring a statistical origin. Third, and most decisively, we measured the projected radial distribution of satellite galaxies, $\Sigma_{\rm sat}$, as a direct, lensing-independent tracer of the underlying mass distribution. The satellite profiles are smooth and continuous at all angular scales and across all BCG stellar mass bins, with no depression at the scales corresponding to the gap. This strongly disfavours a genuine absence of mass as the cause of the signal deficit, and together with the previous two tests, provides strong evidence that the \textit{Einstein Gap} arises from a process affecting gravitational lensing exclusively.

A key observational result that emerged from the mass-binned analysis is that the depth and prominence of the \textit{Einstein Gap} increase systematically with BCG stellar mass, and therefore with total halo mass via the stellar-to-halo mass relation. 
We quantify this statistically, finding the gap detected at 
$N_\sigma>29$ across the highest-mass bins (BIN5–BIN7) and at $N_\sigma\sim13$ for the full ZOU sample, once the full covariance of the stacked profile is taken into account.
This mass-dependent scaling is inconsistent with any methodological or statistical artifact, and constitutes strong independent evidence that the feature is physically real. It also provides a direct observational handle for future studies: the gap properties (its angular position and the sharpness of its boundaries) encode information about the total halo mass and its central concentration that is complementary to what can be extracted from standard weak lensing analyses.

To further test the strong-lensing hypothesis, we developed an improved magnification bias simulator incorporating full ray-tracing via the lens equation and a SISSA mass density profile, which combines an NFW dark matter halo with a Sérsic stellar component. The simulations reproduce the observed signal deficit at $\gtrsim 10''$ for an average cluster halo mass of $\sim 5\times10^{14}\,M_\odot$, in good agreement with the observations. The \textit{Einstein Gap} emerges naturally as a consequence of strong lensing: background sources whose unlensed positions fall within the Einstein radius are displaced outward, creating a characteristic empty annulus in the stacked map bounded by the Einstein ring on the outside and by the inner critical curve on the inside. The angular position of the gap scales as $\theta_E \propto M_{\rm H}^{2/3}$, providing a natural explanation for its mass-dependent prominence. While the simulator reproduces the outer boundary of the gap with high fidelity, discrepancies remain in the inner region, most likely due to the rigidity of the SISSA power-law parametrization and the absence of satellite contributions in the current implementation. These limitations motivate future work with higher-fidelity lensing simulations incorporating the full complexity of the cluster mass distribution.

The \textit{Einstein Gap} therefore represents a genuine strong lensing feature of massive halos that appears consistent with signal deficits visible in weak-lensing profiles published over roughly two decades, though not previously interpreted as a strong-lensing feature. Its identification opens several promising directions for future research. The scaling of the gap with halo mass offers a new, independent method to constrain the mass and concentration of galaxy clusters from magnification bias measurements alone. The comparison between PEAK- and BCG-centred analyses provides a new observational handle on the degree of substructure lensing in massive systems, and could help determine whether the lensing signal within groups and clusters originates from a smooth common halo or from the cumulative contribution of individual member galaxies. A further promising diagnostic would be to stratify clusters by richness independently of BCG stellar mass, in order to test directly whether the cumulative lensing contribution of the satellite population, rather than halo mass alone, regulates the suppression of the \textit{Einstein Gap} in the richest systems. More broadly, the results presented here demonstrate that magnification bias, when combined with sufficiently accurate source positions, probes a lensing regime that is inaccessible to conventional shear-based analyses, and that this regime contains rich physical information about the inner mass distribution of galaxy clusters that has yet to be fully exploited.

However, current facilities cannot meet the requirements for precision magnification bias measurements because none combine high sensitivity with wide-area coverage at sub-millimetre wavelengths. Future facilities capable of combining high astrometric accuracy with broad sky coverage, as the proposed Atacama Large Aperture Submillimeter Telescope \citep[AtLAST,][]{BOO24}, will therefore be key to unlocking the full potential of magnification bias, enabling precise halo studies from kiloparsec scales to the outskirts of galaxy systems.

\appendix
\section{Theoretical framework}
\label{ANX:profiles}
The gravitational lensing effect modifies the integral number counts of background sources in a flux-limited sample due to the presence of a mass distribution between these sources and the observer. This modification appears from the combined effects of magnification, which amplifies the flux of fainter sources, and dilution, which increases the apparent solid angle on the sky. Consequently, the number of background sources per unit solid angle and redshift, with an observed flux density greater than $S$, is altered at each two-dimensional angular position $\vec{\theta}$ on the celestial sphere as \citep{Bar01}.

\begin{equation}
    n(>S,z;\vec{\theta})=\frac{1}{\mu(\vec{\theta})}n_0\Big(>\frac{S}{\mu(\vec{\theta})},z \Big),\label{eq2}
\end{equation}

where $n_0$ represents the integrated number counts in the deficiency of magnification, and $\mu(\vec{\theta})$ denotes the magnification field at the angular position $\vec{\theta}$. Assuming that the unlensed integrated number counts follow a redshift-independent power-law behaviour, given by $n_0(>S,z)=AS^{-\beta}$, Eq. \ref{eq2} can be expressed as

\begin{equation}
\frac{n(>S,z;\vec{\theta})}{n_0(>S,z)}=\mu^{\beta-1}(\vec{\theta}).
\label{eq3}
\end{equation}

The relationship between the magnification field due to a sample of lenses and the cross-correlation observable that we aim to measure becomes evident upon interpreting the physical significance of the above ratio. The left-hand side of equation \ref{eq3} quantifies the excess or deficit of background sources in the direction $ \vec{\theta}$, as seen by a lens at redshift $z_l$, relative to the scenario without lensing. The angular CCF between a foreground sample of lenses at redshift $z_l$ and a background sample of objects at redshift $z_b$ is defined as

\begin{equation}
    w_x(\vec{\theta};z_l,z_b)\equiv\langle\delta n_f(\vec{\phi})\,\delta n_b(\vec{\phi}+\vec{\theta})\rangle,
\end{equation}
where $\delta n_b$ and $\delta n_f$ denote the background and foreground object number density variations, respectively. Since we are stacking the lenses at a fixed position, it follows from the above argument that

\begin{equation}
    w_x(\vec{\theta};z_l,z_b)=\mu^{\beta-1}(\vec{\theta})-1.
\end{equation}

This expression holds under the assumption that magnification is the dominant effect and that the magnified background number counts follow a power-law slope parameterized by $\beta$. It is important to emphasize that this is a general relationship and is not limited to the weak lensing regime, $\mu\ll1$. In the weak limit, $\mu\approx1+2\kappa$, the expression can be expanded, but the full form remains valid as long as the source counts respond uniformly to magnification.

Suppose a lens, located at an angular diameter distance $D_d$ from the observer, deflects the light rays from a source at an angular diameter distance $D_s$. Let $\vec{\theta}=\vec{\xi}/D_d$ denotes the angular position of a point on the image plane. In this context, the convergence field can be defined as

\begin{equation}
\kappa(\vec{\theta})=\frac{\Sigma(D_d\vec{\theta})}{\Sigma_{\text{cr}}},
\end{equation}
where $\Sigma(\vec{\xi})$ denotes the mass density projected onto a plane perpendicular to the light ray and $\Sigma_{\text{cr}}$ is the critical mass density described as
\begin{equation}
    \Sigma_{\text{cr}}=\frac{c^2}{4\pi G}\frac{D_s}{D_dD_{ds}}
\end{equation}
where $D_{ds}$ represents the angular diameter distance from the lens to the background source.\\

If the lens is assumed to be axially symmetric, the origin can be chosen at the symmetry centre, leading to $\kappa(\vec{\theta})=\kappa(\theta)$. In this case, the magnification field, $\mu(\theta)$, is related to the convergence via \citep{Bar01}

\begin{equation}
    \mu(\theta)=\frac{1}{(1-\bar{\kappa}(\theta)(1+\bar{\kappa}(\theta)-2\kappa(\theta))}.
\end{equation}

In this context, $\bar{\kappa}(\theta)$ denotes the mean surface mass density enclosed within the angular radius $\theta$, given by

\begin{equation}
    \bar{\kappa}(\theta)= \dfrac{2}{\theta^2} \cdot\int_{0}^{\theta} d\theta' \theta' \kappa(\theta')
\end{equation}

\subsection{Navarro-Frenk-White profile}
\label{sec:NFW}

 The Navarro-Frenk-White \citep[NFW,][]{NAV96} mass density profile is one of the most used profiles in cosmology. It is a two-parameter model given by
 \begin{equation}
     \rho_{\text{NFW}}(r;r_s,\rho_s)= \frac{\rho_s}{(r/r_s)(1+r/r_s)^2},
 \end{equation}

which can be shown to satisfy \citep{SCH06}

\begin{equation}
    \kappa_{\text{NFW}}(\theta)=\frac{2r_s\rho_s}{\Sigma_{\text{cr}}}f(\theta/\theta_s)\quad\quad\quad \bar{\kappa}_{NFW}(\theta)=\frac{2r_s\rho_s}{\Sigma_{cr}}h(\theta/\theta_s),
\end{equation}
where $\theta_s\equiv r_s/D_d$ is the angular scale radius,
\begin{equation}
    f(x)\equiv \begin{cases} \frac{1}{x^2-1}-\frac{\arccos{(1/x)}}{(x^2-1)^{3/2}}\quad\quad&\text{if } x>1\\
    \,\,\frac{1}{3} \quad\quad&\text{if } x=1\\
    \frac{1}{x^2-1}+\frac{\text{arccosh}(1/x)}{(1-x^2)^{3/2}} \quad\quad&\text{if } x<1
    \end{cases}
\end{equation}
and
\begin{equation}
    h(x)\equiv \begin{cases} \frac{2}{x^2}\Big(\frac{\arccos{(1/x)}}{(x^2-1)^{1/2}}+\log{\frac{x}{2}}\Big)\quad\quad&\text{if } x>1\\
    \,{\scriptstyle 2\,(1-\log{2})} \quad\quad&\text{if } x=1\\
    \frac{2}{x^2}\Big(\frac{\text{arccosh }(1/x)}{(1-x^2)^{1/2}}+\log{\frac{x}{2}}\Big) \quad\quad&\text{if } x<1
    \end{cases}.
\end{equation}

\subsection{SISSA profile}
\label{sec:SISSA}

The SISSA model \citep{LAP12} describes the mass distribution of a lens galaxy using a two-component approach: a stellar component modelled with a Sérsic profile and a dark matter halo represented by NFW profile. Given the halo mass $M_{h}$ (and the virialization redshift), the model is fully characterized by three parameters: the halo-to-stellar mass ratio $M_{h}/M_\star$, the Sérsic index $n$, and the halo concentration parameter, which can be derived from a mean mass–concentration relation \( c(M, z) \).

As detailed in \cite{LAP12}, the combined stellar and dark matter contributions to the total surface mass density can be approximated by a power law within the radial range $-2.5 \lesssim \langle\log(s/R_{h})\rangle \lesssim -1$, where $s$ denotes the projected radial coordinate on the plane of the sky and $R_{h}$ is the halo virial radius. This region generally dominates the gravitational deflection. In this region the surface density is expressed as
\begin{equation}\label{eq:power_law}
\Sigma(s)=\Sigma_0\,\left({\dfrac{s}{s_0}}\right)^{-\eta}.
\end{equation}
where $\Sigma_0$ is the normalization at the reference radius $s_0\approx 10^{-2}\, R_{\rm H}$, and $\eta$ is the power-law index. According to \cite{LAP12}, for a fixed halo mass, both $\Sigma_0$ and $\eta$ (typically in the range ($0.8-0.9$) show only weak dependence on the mass distribution parameters (see their Table 1 for further details).

Under this power-law approximation, the lens equation simplifies significantly, allowing for faster computations of lensing properties. The deflection angle within a circular aperture of angular radius $\theta$ is given by
\begin{equation}
\bar \alpha(\theta)=|\bar \theta|^{1-\eta}.
\end{equation}
where the over-bar notation (e.g., $\bar \theta\equiv
\theta/\theta_E$) denotes normalization to the Einstein radius $\theta_E$. For power-law mass profiles, the Einstein radius can be expressed as
\begin{equation}\label{eq:thetaE}
\theta_E=\theta_0\, \left({\dfrac{2}{2-\eta}}\, {\dfrac{\Sigma_0}
{\Sigma_c}}\right)^{1/\eta},
\end{equation}
with $\theta_0\equiv s_0\,(1+z_\ell)/D_\ell$, where $z_\ell$ is the lens redshift, $D_\ell$ is the angular diameter distance to the lens, and $\Sigma_c$ is the critical surface mass density for lensing.

The magnification $\mu$ of an image is given by
\begin{equation}\label{eq:mu}
\mu={\dfrac{1}{
[1-|\bar\theta|^{-\eta}]\,[1-(1-\eta)\,|\bar\theta|^{-\eta}]}}.
\end{equation}
This expression reveals the presence of two critical curves: the standard one at $\theta=\theta_E$ (corresponding to the Einstein ring) and an inner critical curve that emerges when $\eta<1$, located at
\begin{equation}
\label{eq:inner_ring}
\theta_I=\theta_E\, (1-\eta)^{1/\eta}.
\end{equation}
To compute the magnification from Eq.~\eqref{eq:mu}, the image positions $\bar\theta$ must be determined as functions of the source’s normalized angular separation $\bar{\beta} = \beta / \theta_E$. This is achieved by solving the lens equation:
\begin{equation}
\bar \beta=\bar \theta\, -{\dfrac{\bar\theta}{
|\bar\theta|}}\,|\bar\theta|^{1-\eta}.
\end{equation}
Summing the contributions from all images provides the total magnification as a function of $\beta$.

\section{Additional robustness tests}
\label{ANX:Validation}
\subsection{Validation of the Analytical Estimation for the Random Signal}

\begin{figure*}
\centering
\includegraphics[width=1\textwidth]{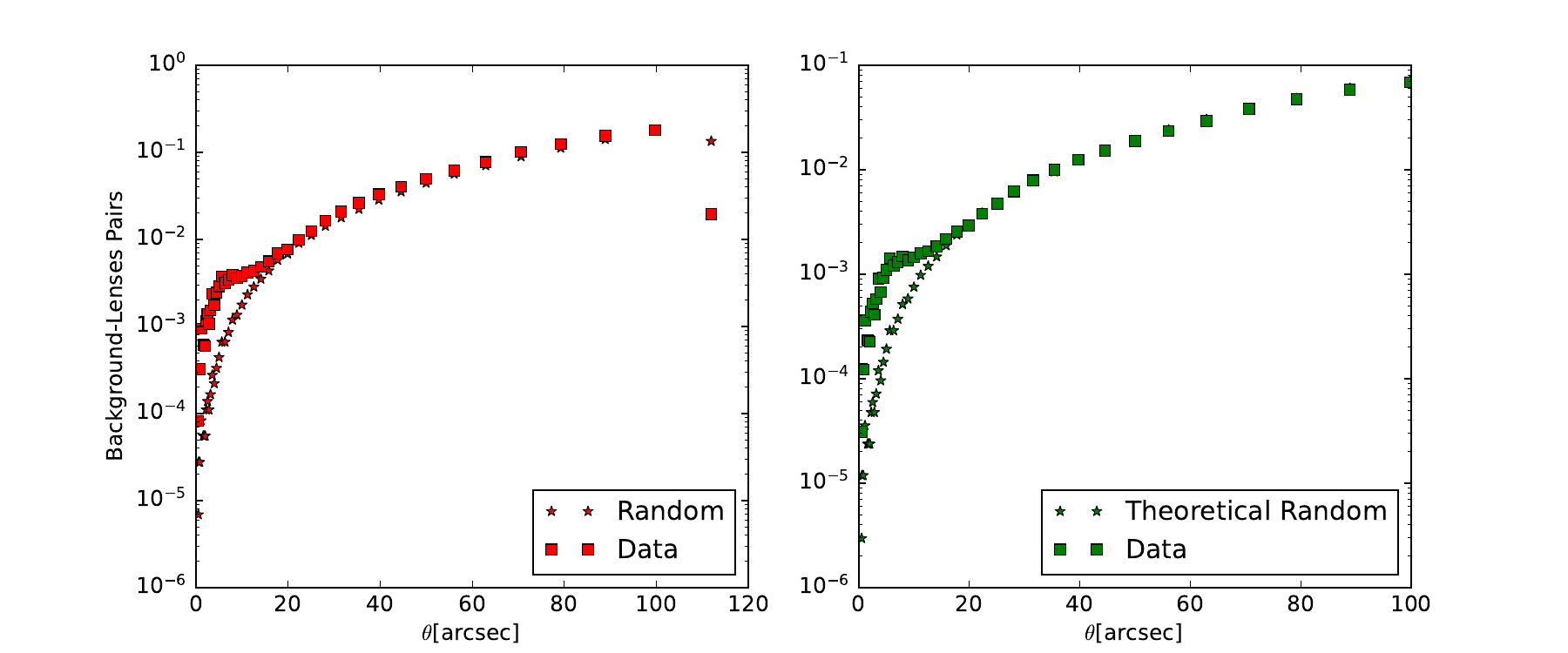}
\caption{Left panel: comparison between the stacked data profile (red circles) and the result obtained using a simulated random catalogue (red stars). Right panel: same data profile (green circles) compared to the analytical $RR(\theta)$ estimate (green stars). Both methods yield consistent results.}
\label{fig.2.19}
\end{figure*}

To verify the robustness of our analysis against possible biases in the estimation of the random signal, we compare two different approaches for computing the $RR(\theta)$ term, which enters the cross-correlation function estimator.

The first method relies on a simulated random catalogue. Following standard practice, we generate a mock sample of background sources with random positions, typically containing ten times the number of foreground lenses. These random sources are stacked around the real lens positions using the same procedure as for the data, yielding a random stacked map (left panel in Fig.~\ref{fig.2.19}). This approach accounts for the survey geometry and selection effects, ensuring consistency with the observational setup.

The second method, introduced in \cite{CRE22}, is an analytical estimation that assumes a uniform background source density $\rho$. Under this assumption, the expected number of random lens–source pairs in an annulus of radius $\theta$ and width $\Delta\theta$ is given by:
\begin{equation}
RR(\theta) = \rho \cdot A_{\text{ring}}(\theta),
\end{equation}
where the annular area can be approximated as:
\begin{equation}
A_{\text{ring}}(\theta) \approx 2\pi\theta\Delta\theta,
\end{equation}
leading to the final expression:
\begin{equation}
RR(\theta) = \rho \cdot 2\pi\theta\Delta\theta.
\end{equation}

This analytical model provides a fast and smooth estimate of the random signal without requiring the generation of artificial catalogues, which can be advantageous in large-scale studies or when testing different configurations.

We performed a full circular analysis using both approaches, with radial binning and Jackknife resampling to compute uncertainties. The comparison, shown in Fig.~\ref{fig.2.19}, demonstrates excellent agreement between the two methods across the full range of scales considered in this study. This demonstrates that, for our dataset and analysis window (extending up to $\sim$1 Mpc), the theoretical approximation is valid and does not introduce systematic errors.

We emphasize, however, that for analyses targeting larger scales or datasets with complex masking, random catalogs may be preferred to capture survey boundaries and angular incompleteness. 

\subsection{Comparison of cross-matching strategies}
\label{ANX:xmatch}

\begin{figure*}
\centering
\includegraphics[width=0.49\linewidth]{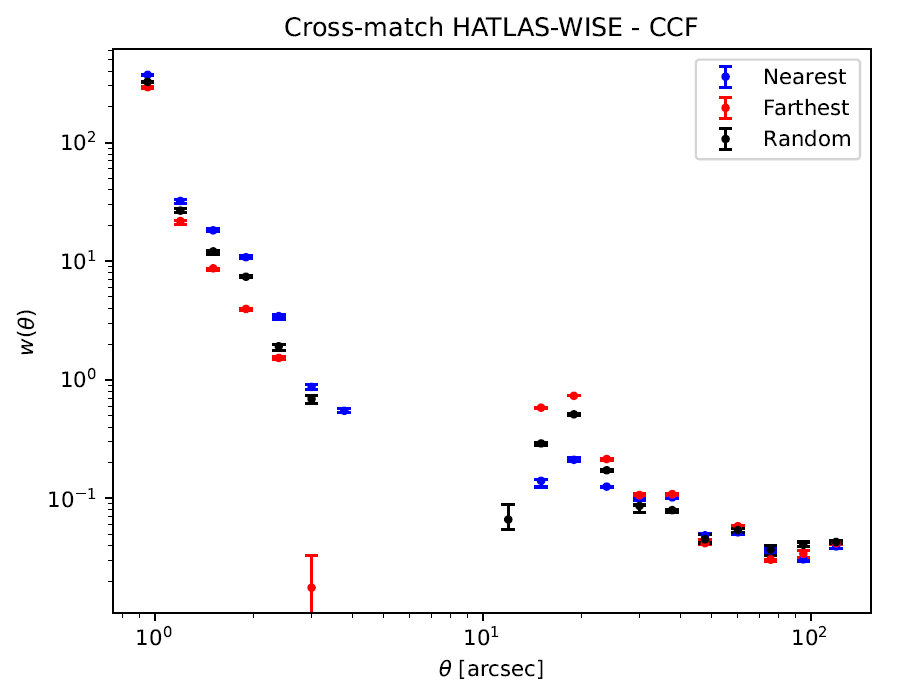}
\includegraphics[width=0.49\linewidth]{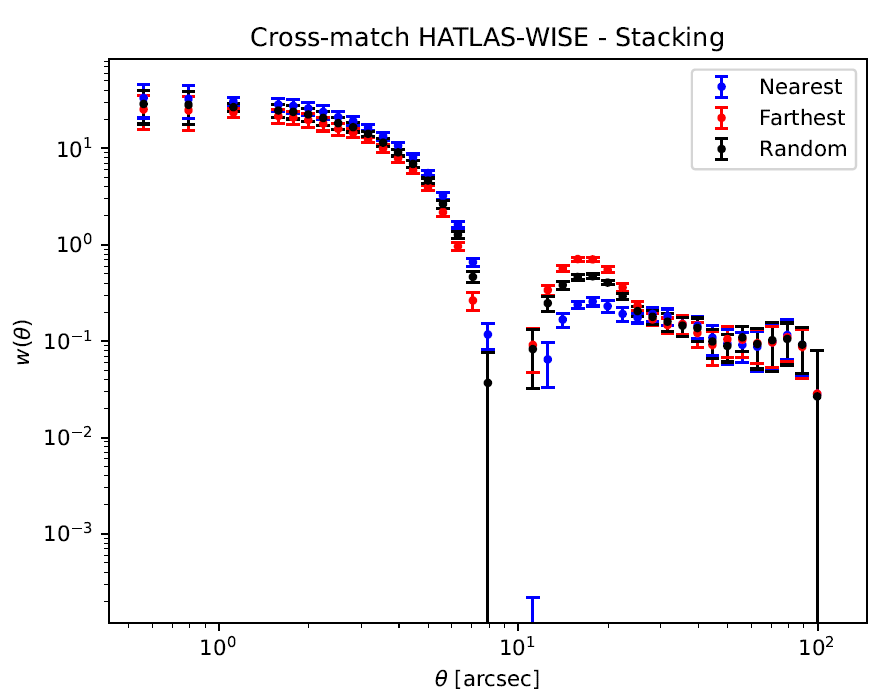}
\caption{Comparison between different cross-matching strategies (left: traditional cross-correlation; right: stacking estimator), using nearest (red), farthest (blue), and random (black) counterparts within $17.6''$. The \textit{Einstein Gap} is present in all cases.}
\label{fig:xmatc}
\end{figure*}

To validate the robustness of the observed signal drop against methodological biases, we compare the magnification signal obtained using two estimators (stacking and cross-correlation) and three different cross-matching strategies between the H-ATLAS and WISE catalogues: selecting the nearest, farthest, or a random infrared source within the $17.6''$ matching radius.

As shown in Figure~\ref{fig:xmatc}, the \textit{Einstein Gap} feature persists across all configurations, indicating that it is not an artifact of the matching criterion or estimator used. The clear difference at the smallest angular scales between both methodologies is due to the fact that the traditional CCF cannot take into account the positional uncertainties producing a more concentrated distribution due to the lensing probability. This is an example of the advantage of using the stacking technique to study the central part of halos.

\subsection{Dependence of the \textit{Einstein Gap} on the positional smoothing kernel}
\label{ANX:smooth}

\begin{figure}
\centering
\includegraphics[width=\linewidth]{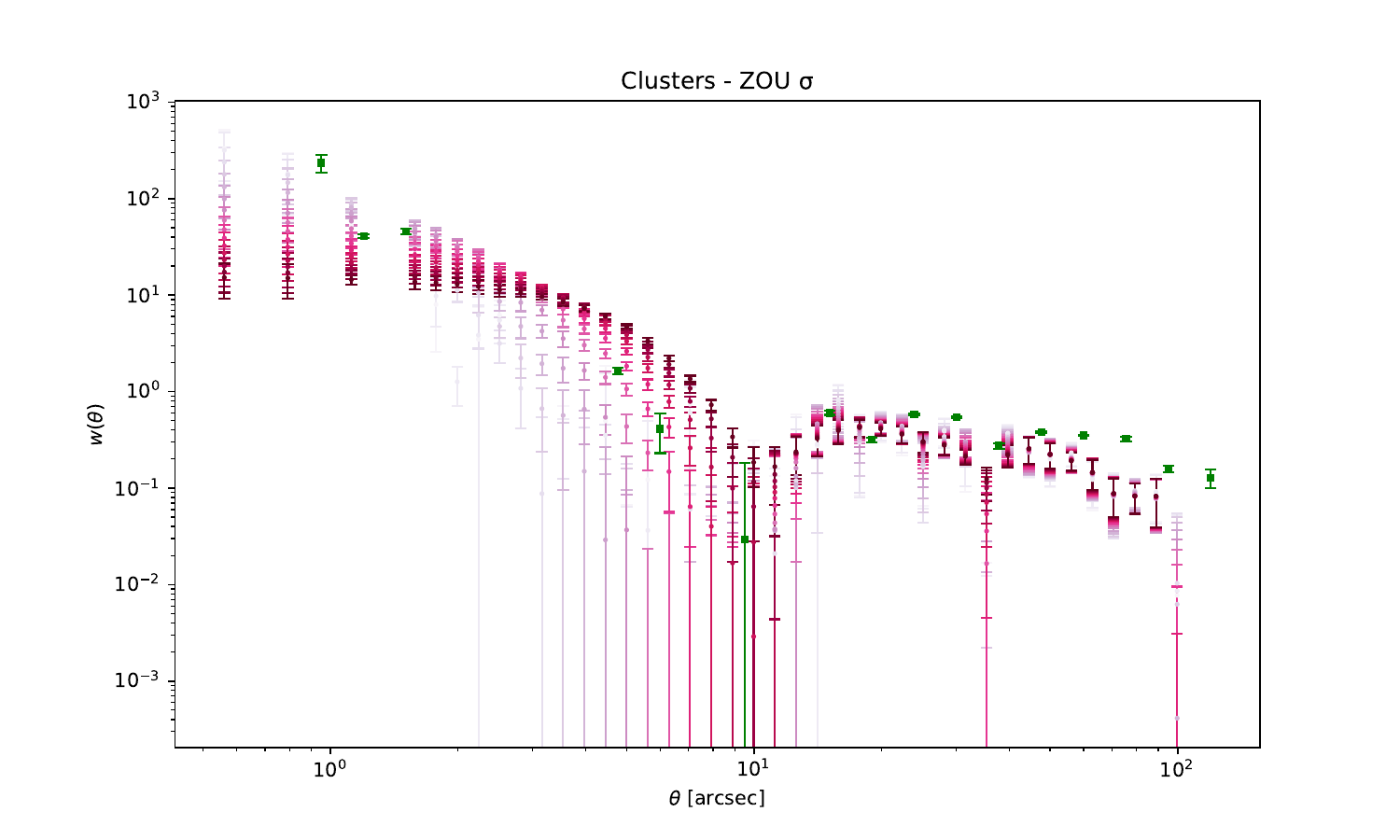}
\caption{Stacking measurements of the CCF between ZOU clusters and H-ATLAS/WISE background sources for smoothing kernels of width between $0''$ and $3''$ in steps of $0.2''$ (magenta points, from lighter to darker with increasing kernel width), compared with the traditional Landy-Szalay estimator (green points). The \textit{Einstein Gap} becomes progressively wider and better defined as the smoothing is reduced, demonstrating that the feature is not a smoothing artifact but a genuine physical signal whose full extent is only revealed at the highest angular resolution achievable with our data.}
\label{fig:s0_to_s3}
\end{figure}

Figure~\ref{fig:s0_to_s3} shows the stacking measurements obtained using smoothing kernels of width between $0''$ and $3''$ in steps of $0.2''$ (from lighter to darker magenta), together with the traditional Landy-Szalay estimator (green points) for reference. As the smoothing kernel is reduced, the \textit{Einstein Gap} becomes progressively wider and better defined, with its boundaries sharpening toward both smaller and larger angular separations. This behaviour is the opposite of what would be expected from a systematic artifact or a statistical fluctuation, both of which would become less coherent as the noise level increases with reduced smoothing. The fact that reducing the smoothing reveals a more prominent and better-defined gap is therefore a clear signature of a genuine physical feature in the data, providing an independent validation of the conclusions drawn in Section~\ref{sec:ruling_out}.

\subsection{Mass density profile analysis using BCG positions as centres} 
\label{ANX:BCG_centre} 
\begin{figure}
    \centering
    \includegraphics[width=0.87\textwidth]{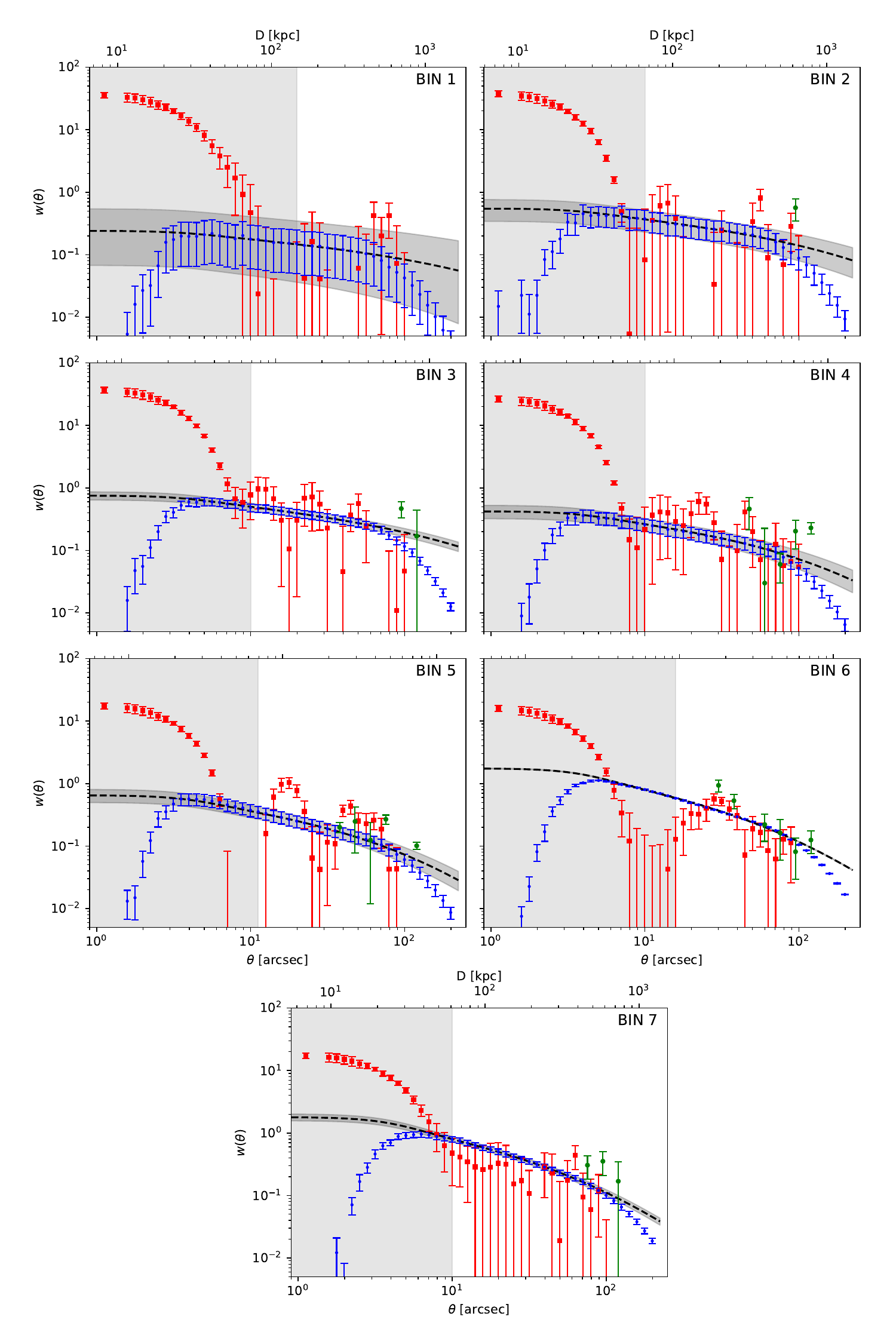}
\caption{Mass density profile analysis for each mass bin (BCG centring). The panels are ordered by BCG stellar mass range from top to bottom and left to right, BIN1 being at the top left and BIN7 being the panel at the bottom. The red circles represent the results obtained using the lensing method, the green ones using the traditional estimator, while the blue squares correspond to the renormalised $\Sigma_{\rm sat}$ points. The best NFW fit to the outer region of the satellite number density profile for each BCG stellar mass range is represented by a dashed black line, and the shaded black region indicates its 68\% uncertainty. Measurements within the gray areas were discarded for this analysis. }
    \label{fig.BINmasa}
\end{figure}

Figure~\ref{fig.BINmasa} presents the mass density profile analysis for each BCG stellar mass bin, equivalent to Fig.~\ref{fig.BINmasa_PEAK} but using the BCG position as the cluster centre rather than the density peak. As discussed in Section~\ref{sec:miscenter}, the choice of centre has a significant impact on the detectability of the \textit{Einstein Gap}, particularly in the highest mass bins where the BCG is more centrally located within the cluster and its own strong lensing contribution becomes relevant. Comparing both figures reveals that the PEAK-centred analysis yields a cleaner and more regular \textit{Einstein Gap} whose prominence increases more uniformly with BCG stellar mass. However, the BCG-centred results produce smoother profiles in the highest mass bins where the cumulative satellite contribution fills in the intermediate angular scales where the gap would otherwise emerge. This comparison therefore provides direct observational evidence that the internal mass distribution of the lensing system, and not just its total mass, governs the detectability of the \textit{Einstein Gap}.

\section{Miscentring Figures}
\label{ANX:figures}

\begin{figure}
    \centering
\includegraphics[width=0.4\textwidth]{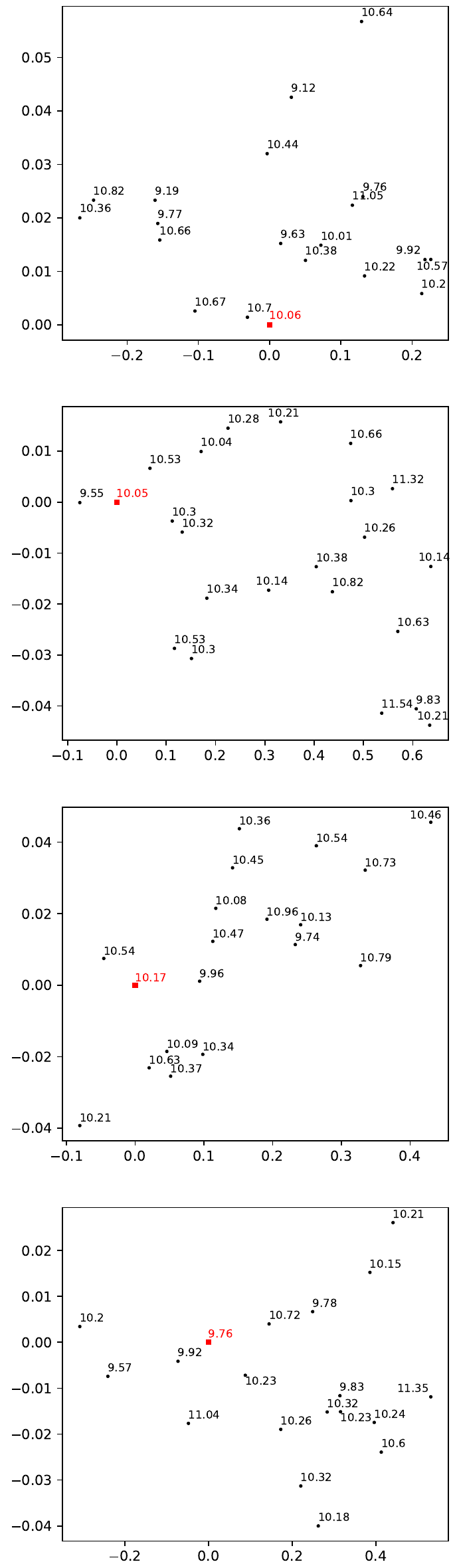}
\includegraphics[width=0.4\textwidth]{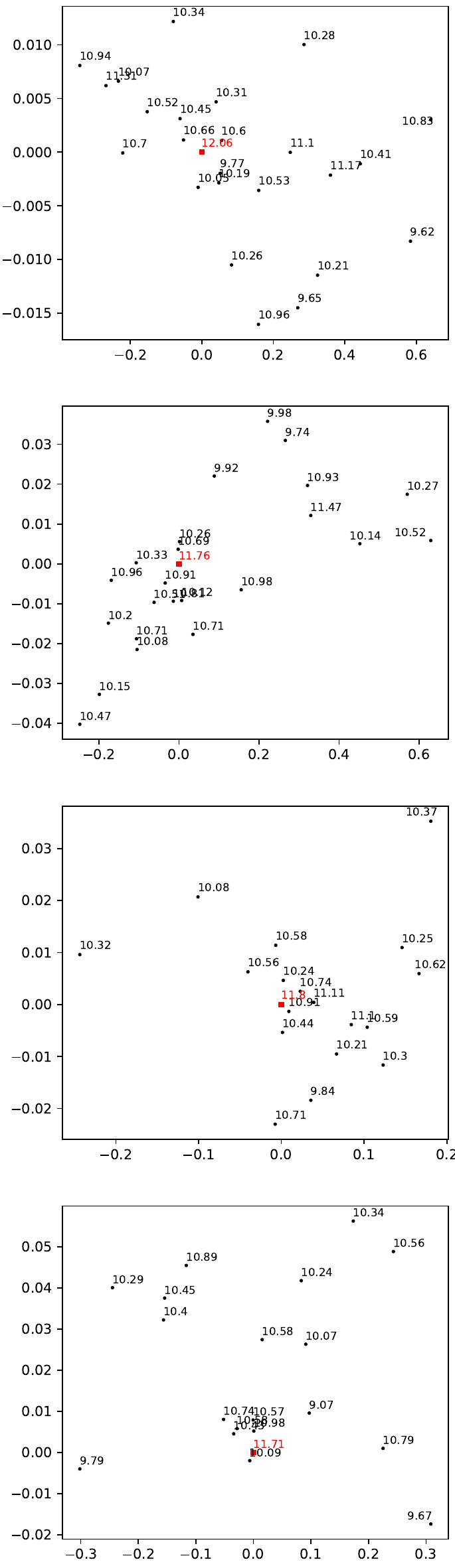}
    \caption{ Relative positions of satellites with respect to the BCG centre (marked in red) for four randomly selected clusters in BIN1 (left column) and for another four randomly selected clusters in BIN7 (right column). The logarithmic mass in $M_\odot$ of each galaxy is displayed next to its corresponding point.The axis units are in degrees.}
    \label{fig.BIN1vsBIN7posv2}
\end{figure}

Figure \ref{fig.BIN1vsBIN7posv2} illustrate the relative positions of satellite galaxies with respect to the BCG centre (marked in red) for Bin 1 and Bin 7. This selection highlights the contrast between the two mass extremes. In Bin 1, BCGs do not appear to be centrally located within their clusters, and their stellar mass is often lower than that of many satellite galaxies. Instead, in Bin 7, BCGs tend to be more centrally positioned, with their stellar mass exceeding that of most satellites. The decision to focus on these two bins rather than displaying all seven was made to better emphasise this trend, as the transition appears more gradual when all bins are included.

\begin{figure}
    \centering
    \includegraphics[width=0.9\textwidth]{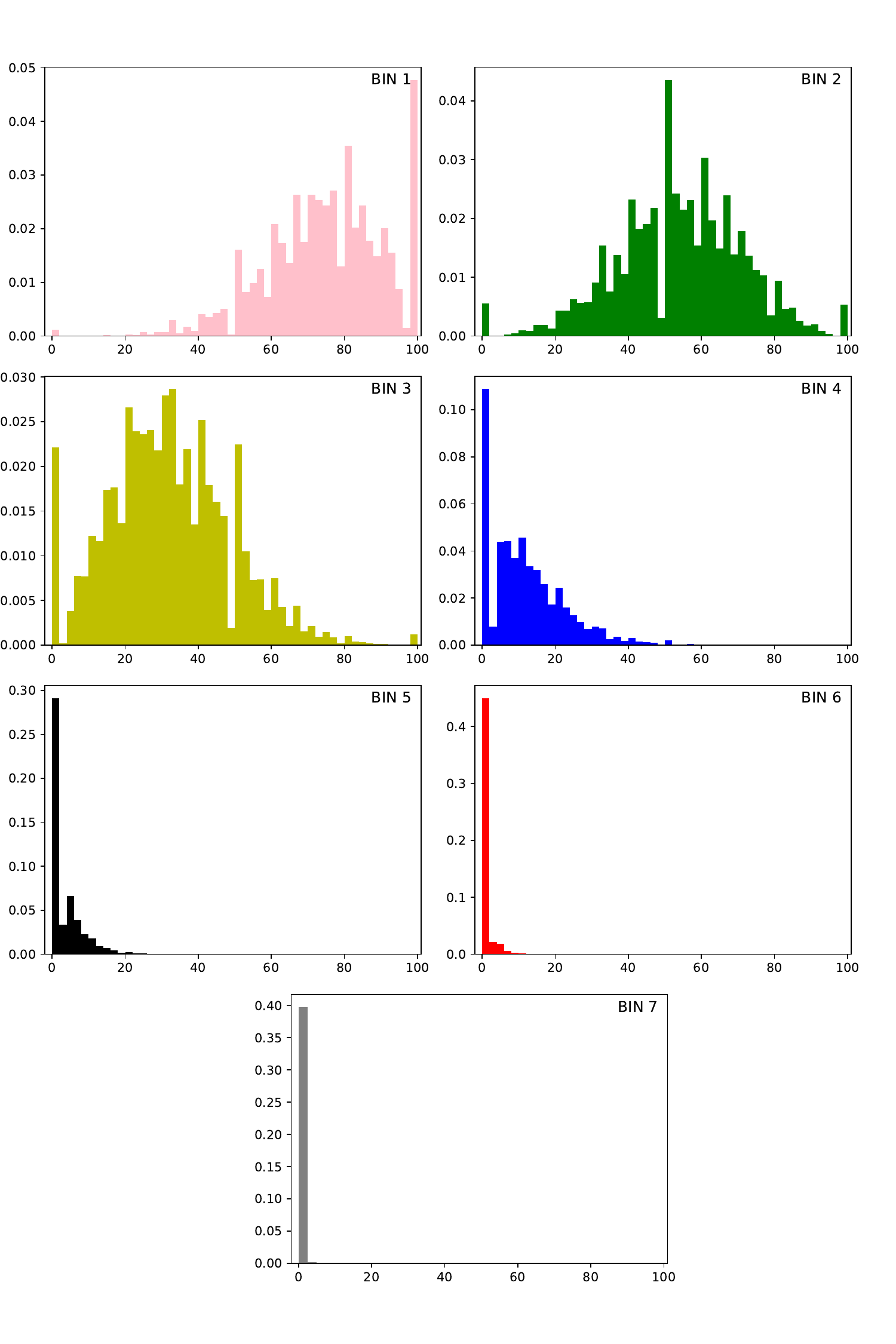}

    \caption{Histograms showing the number of galaxy clusters with a percentage of satellites with a mass greater than their BCG for each mass range described in Table \ref{tableMass}.}
    \label{fig.BINpercent}
\end{figure}

Figure \ref{fig.BINpercent} presents histograms showing the number of galaxy clusters with a percentage of satellite galaxies that exceed the stellar mass of their respective BCG for each mass ranges defined in Table \ref{tableMass}. This figure allows us to see a clear trend: the lower the BCG stellar mass interval the higher percentage of satellites have a stellar mass greater than the stellar mass of their BCG.

\acknowledgments

DC, JGN, LB, FFR, JMC acknowledge the PID2021-125630NB-I00 and PID2025-169168NB-I00 projects funded by \\MCIN/AEI/10.13039/501100011033/FEDER, UE. 
LB, JMC also acknowledges the CNS2022-135748 project funded by MCIN/AEI/10.13039/501100011033 and by the EU “NextGenerationEU/PRTR”.
HZ acknowledges  the supports from the National Key R\&D Program of China (grant No. 2023YFA1607800) and National Natural Science Foundation of China (NSFC; grant Nos. 12373010 and 12120101003).

The \textit{Herschel}-ATLAS is a project with \textit{Herschel}, which is an ESA space observatory with science instruments provided by European-led Principal Investigator consortia and with important participation from NASA. The H-ATLAS web-site is http://www.h-atlas.org. GAMA is a joint European-Australasian project based around a spectroscopic campaign using the Anglo-Australian Telescope.

This research has made use of the python packages \texttt{ipython} \citep{ipython}, \texttt{matplotlib} \citep{matplotlib} and \texttt{Scipy} \citep{scipy}.

\bibliographystyle{JHEP}
\bibliography{biblio.bib}
\end{document}